\documentclass[pra,aps,twocolumn,,showpacs,floatfix]{revtex4}

\usepackage{amsmath,amssymb,mathrsfs}
\usepackage{graphicx}
\usepackage{version}
\DeclareGraphicsExtensions{.eps,.ps}

\begin{document}

\title
{Superfluidity without symmetry-breaking: the time-dependent
Hartree-Fock approximation for Bose-condensed Systems}
\author{C.-H. Zhang}
\affiliation{Department of Physics, Indiana University, 727 E.\
3rd Street, Bloomington, IN 47405}
\author{H.A. Fertig}
\affiliation{Department of Physics, Indiana University, 727 E.\
3rd Street, Bloomington, IN 47405}
\date{\today}

\begin{abstract}
We develop a time-dependent Hartree-Fock approximation that is
appropriate for Bose-condensed systems. Defining a {\it depletion
Green's function} allows the construction of condensate and
depletion particle densities from eigenstates of a single
time-dependent Hamiltonian, guaranteeing that our approach is a
conserving approximation. The poles of this Green's function yield
the energies of number-changing excitations for which the condensate
particle number is held fixed, which we show has a gapped spectrum
in the superfluid state. The linearized time-dependent version of
this has poles at the collective frequencies of the system, yielding
the expected zero sound mode for a uniform infinite system. We show
how the approximations may be expressed in a general linear response
formalism.

\end{abstract}


\maketitle


\section{Introduction}

The time-dependent Hartree-Fock approximation (TDHFA) is a powerful
method for determining the properties of many-body systems\
\cite{Blaizot86}.  It allows the computation of the density response
functions, conductivities, and other linear response functions.  The
poles of such functions determine the collective modes of a system,
and within the TDHFA the effects of thermally-generated single
particle excitations on these modes are included. The TDHFA has thus
become a standard tool in condensed matter, nuclear, and atomic
physics.

Surprisingly, the TDHFA has not enjoyed such success in the area of
Bose-condensed systems\ \cite{giorgini}.  In part this is because
there are other powerful methods for computing collective modes,
based on the Bogoliubov approximation in which there is coherence
among states with different particle numbers; i.e., where gauge
symmetry is spontaneously broken. Field theoretic techniques that
incorporate this were developed long ago,\ \cite{old} but in light
of the significant progress on atomic gases in the last decade\
\cite{Pethick02}, there is much renewed effort to improve upon these
methods\ \cite{Griffin96}. Among these are generalizations of
various mean-field schemes to properly include the effects of
symmetry-breaking\ \cite{tosi}, and a ``dielectric approach'' that
carefully treats Green's functions and response functions in a
consistent way\ \cite{Griffin93,dielectric}. The challenge in these
studies has been to treat the dynamics of the condensate and of the
thermally excited (i.e., depleted) particles in a fully consistent
manner.  The resulting schemes are considerably more involved than
the original Bogoliubov approach. (For a review, see Ref.\
\onlinecite{Griffin96}.)

As we show below, one way to avoid the difficulties of disentangling
the single-particle and collective excitations is to work with an
ensemble in which the gauge symmetry is {\it not} broken\
\cite{fix_n}. In situations for which the particle number is
relatively small, this may actually be preferable to broken symmetry
approaches for which the associated fluctuations in the particle
number can become problematic. Our chosen approach is the TDHFA. Its
principle advantage is that the condensate wavefunction and the
single-particle excitations appear as states of a {\it single}
time-dependent, non-local Hamiltonian, so that they may be treated
on an equal footing with exchange effects fully included. The
connection through a single Hamiltonian guarantees that conservation
laws will be respected\ \cite{Kadanoff62}, and we will demonstrate
that the method correctly produces the gapless superfluid mode for a
homogenous system.

Our principal conclusions are as follows. (1) We find that in order
to properly deal with exchange, one must adopt a {\it constrained}
grand-canonical ensemble for the density matrix first introduced by
Huse and Siggia\ \cite{Huse82}, in which the number of particles in
the condensate is fixed while the occupation probability for other
single particle states is given by the standard grand canonical
ensemble. Within this ensemble, one finds a {\it gap} in the single
particle spectrum between the condensate and the other single
particle states\ \cite{Huse82,others}, the latter of which we call
depletion states\ \cite{depletion}. This exchange gap is analogous
to the single particle gap that arises in superconducting systems
and only arises when the system is Bose-condensed, and thus may be
viewed as a superfluid order parameter. The gap should be observable
in a tunneling experiment between a normal Bose gas and a Bose
condensate. (2) We define a {\it depletion Green's function}
$\tilde{G}({\bf q},\omega)$, whose poles occur at the
single-particle energies of the depletion states in the Hartree-Fock
approximation, which by construction does {\it not} have a pole at
the collective mode frequency, as is expected for the full Green's
function. The response of this Green's function to a weak,
time-dependent perturbation, coupled with the equation of motion for
the condensate wavefunction -- which is a finite temperature
generalization of the GP equation -- allows us to generate equations
of motion for the condensate and depletion states governed by the
{\it same} effective Hamiltonian, and to define response functions
whose poles occur at the collective excitations of the system. This
naturally captures the interplay between the condensate and
depletion states. (3) We solve these coupled equations for the
simplest case of an infinite uniform Bose gas, and demonstrate that
whenever there is Bose condensation, there is a gapless collective
(zero sound) mode.  This is usually identified as the superfluid
mode in approaches where the gauge symmetry is broken; in our
approach we find the mode even though the symmetry is kept intact.
We demonstrate that the density response function contains structure
that could not arise in simple Bogoliubov approaches where only the
collective mode is retained.

This paper is organized as followings: In Sec.\ \ref{sec:HF-TDHF},
we derive the static and time-dependent HF equations at finite
temperature by using a constrained grand canonical ensemble. We then
show how this can be used to define the depletion Green's function
which captures the energetics of particles outside the condensate.
In Sec.\ \ref{sec:chi}, we demonstrate how the depletion Green's
function may be used to generate an approximate form for response
functions.  We also show that the same result may be obtained
directly from wavefunctions, firmly establishing the connection with
the TDFHA. In Sec.\ \ref{sec:homogenous}, we apply the general
results in previous sections to a homogenous system with a contact
two-body interaction. Discussions and conclusions are presented in
the final section. Further details are presented in the Appendices.

\section{Static and Time-Dependent Hartree-Fock Equations
at Finite Temperature}
\label{sec:HF-TDHF}

We begin with Bose particles in an external potential $U(\vec{r})$
that is time-independent. The Hamiltonian is
\begin{align}
\label{eq:Hamiltonian} \hat{H}&=\int\!\!
d\vec{r}\hat{\psi}^\dagger(\vec{r})H_0 \hat{\psi}(\vec{r})
\nonumber\\
&+\frac12\int\!\!d\vec{r}_1
d\vec{r}_2\hat{\psi}^\dagger(\vec{r}_1)\hat{\psi}^\dagger
(\vec{r}_2)V(\vec{r}_1-\vec{r}_2)\hat{\psi}(\vec{r}_2)
\hat{\psi}(\vec{r}_1),
\end{align}
where $H_0=-\frac{\hbar^2\nabla^2}{2m}+U(\vec{r})$ is the
non-interacting Hamiltonian and $V(\vec{r}_1-\vec{r}_2)$ the
two-body potential. For a neutral Bose gas, $V$ is usually
short-range and can be taken to have a contact form
$V(\vec{r})=g\delta(\vec{r})$ if the gas is dilute. We wish first to
find eigenstates of $\hat{H}$ in the Hartree-Fock approximation
(HFA) at finite temperature\ \cite{Huse82,others}.

We begin with the standard HFA, which we shall see is fine for high
temperatures, but becomes a poor approximation when the system is
Bose-condensed. We seek a single-particle Hamiltonian
\begin{equation}
\label{eq:H_HF} \hat{H}_{HF}=\sum_{\alpha\beta}
\varepsilon_{\alpha\beta} a^\dagger_\alpha a_\beta,
\end{equation}
where the indices $\alpha$ label states of a single-particle basis,
which minimizes the free energy. In terms of these states the
Hamiltonian Eq.\ (\ref{eq:Hamiltonian}) can be written as
\begin{align}
\label{eq:Hamiltonian-in-HF-basis}
H&=\sum_{\alpha\beta}\varepsilon^0_{\alpha\beta}a^\dagger_\alpha
a_\beta +\frac12\sum_{\alpha\beta\delta\gamma}
V_{\alpha\beta\delta\gamma} a^\dagger_\alpha a^\dagger_\beta
a_\delta a_\gamma,
\end{align}
where
\begin{align}
\label{eq:bare-single-particle-matrix}
\varepsilon^0_{\alpha\beta}&=\int\!\!
d\vec{r}\psi^*_\alpha(\vec{r})\left[-\frac{\hbar^2\nabla^2}{2m}
+U(\vec{r})\right]\psi_\beta(\vec{r})
\end{align}
is the matrix element of the non-interacting Hamiltonian, with
$\psi_{\alpha}$ the single particle states,  and
\begin{align}
\label{eq:two-body-interaction-matrix}
V_{\alpha\beta\delta\gamma}=\int\!\!d\vec{r}_1\!\!\int\!\!d\vec{r}_2
\psi^*_\alpha(\vec{r}_1)\psi^*_\beta(\vec{r}_2)
V(\vec{r}_1-\vec{r}_2)\psi_\delta(\vec{r}_2)\psi_\gamma(\vec{r}_1)
\end{align}
is the matrix element of the two-body potential.

At finite temperature $k_BT=1/\beta$ where $k_B$ is the Boltzmann
constant, the expectation value of an operator $\hat{O}$ is
\begin{equation}
\langle\hat{O}\rangle=\mbox{Tr}\hat{D}\hat{O},
\end{equation}
where $\hat{D}$ is the exact density matrix operator.
In the grand canonical ensemble this is
\begin{equation}
\hat{D}=\frac{e^{-\beta(\hat{H}-\mu\hat{N})}}{\mbox{Tr}e^{-\beta
(\hat{H}-\mu\hat{N})}} =\frac{1}{Z}e^{-\beta(\hat{H}-\mu\hat{N})},
\end{equation}
where $Z$ is the corresponding partition function
\begin{equation}
Z=\mbox{Tr}e^{-\beta(\hat{H}-\mu\hat{N})}.
\end{equation}
The Hartree-Fock approximation is based on the variational
principle\ \cite{Blaizot86} that for any trial density matrix
$\hat{D}_r$, one always has for the free energy
\begin{equation}
\Omega[\hat{D}_r]\ge\Omega[\hat{D}].
\end{equation}
In practice
one chooses a form for $\hat{D}_r$ that
minimizes the free energy while allowing
calculations with it to be tractable.
In the Hartree-Fock approximation
we choose 
the trial density matrix
\begin{equation}
\hat{D}_{var}=\frac{1}{Z_{var}}e^{-\beta(\hat{H}_{HF}-\mu\hat{N})}
\end{equation}
with
\begin{equation}
Z_{var}=\mbox{Tr}e^{-\beta(\hat{H}_{HF}-\mu\hat{N})}.
\end{equation}
The parameters
$\varepsilon_{\alpha\beta}$ and $\psi_\alpha$ are
determined by minimizing the
trial free energy \cite{Blaizot86}
\begin{equation}
\Omega_{var}=-\frac{1}{\beta}\ln
Z_{var}-\mbox{Tr}\hat{D}_{var}\hat{H}_{var}
+\mbox{Tr}\hat{D}_{var}\hat{H}.
\end{equation}

Defining the single-particle density matrix
$\rho_{\alpha\beta}$ as
\begin{equation}
\rho_{\alpha\beta}=\mbox{Tr}D_{var}a^\dagger_\alpha a_\beta,
\end{equation}
one has
\begin{equation}
\mbox{Tr}\hat{D}_{var}\hat{H}_{HF} =\mbox{Tr}\rho\varepsilon.
\end{equation}
The variation of $\Omega_{var}$ yields
\begin{align}
\delta\Omega_{var}&=\frac{1}{Z_{var}}\mbox{Tr}e^{-\beta\hat{H}_{var}}
\delta\hat{H}_{var}-\mbox{Tr}\delta\rho\varepsilon-\mbox{Tr}\rho\delta
\varepsilon+\delta\langle\hat{H}\rangle
\nonumber\\
&=-\mbox{Tr}\varepsilon\delta\rho+\delta\langle
\hat{H}\rangle=\mbox{Tr}\left(\frac{\delta\langle\hat{H}\rangle}{\delta\rho}
-\varepsilon\right)\delta\rho,
\end{align}
with
\begin{align}
\langle\hat{H}\rangle
=\sum_{\alpha\beta}\varepsilon^0_{\alpha\beta}\rho_{\alpha\beta}
+\frac12\sum_{\alpha\beta\delta\gamma}V_{\alpha\beta\gamma\delta}
\left[\rho_{\alpha\gamma}\rho_{\beta\delta}+\rho_{\alpha\delta}
\rho_{\beta\gamma}\right].
\end{align}
By requiring $\delta\Omega_{var}=0$, one arrives at
\begin{equation}
\varepsilon_{\alpha\beta}
=\frac{\partial\langle\hat{H}\rangle}{\partial\rho_{\beta\alpha}}
=\varepsilon^0_{\alpha\beta}+\sum_{\gamma\delta}
\bar{V}_{\alpha\gamma\beta\delta}\rho_{\delta\gamma},
\label{epsab}
\end{equation}
where
$\bar{V}_{\alpha\gamma\beta\delta}=V_{\alpha\gamma\beta\delta}
+V_{\alpha\gamma\delta\beta}$. This is the Hartree-Fock equation
in an arbitrary single-particle basis.

\subsection{Constrained Grand Canonical Ensemble}

The above derivation is valid for all temperatures,
and the condensate plays no special
role. However, the result has significant
problems for a Bose-condensed system:
(1) It predicts
macroscopically large fluctuations in the total particle number of
the system below the critical temperature, which is unphysical. In
the diagonal basis where
$n_\alpha=\rho_{\alpha\alpha}=(e^{\beta(\varepsilon_\alpha-\mu)})^{-1}$,
the source of this problem may be traced to the parameter
$\varepsilon_{\alpha=0}-\mu$ associated with the condensate
wavefunction, which becomes arbitrarily small in the thermodynamic
limit. (ii) With a contact interaction, the grand canonical ensemble
produces the same mean-field potential
for all particles, and does not yield
the expected appearance of an exchange energy
only when particles are
in different states, not when they are in the same state. This
is not a serious problem when the system is above the critical
temperature since each level is microscopically occupied. However,
this is a poor approximation when the system is Bose
condensed.

To correct these problems, one may introduce a {\it constrained}
trial grand canonical ensemble\ \cite{Huse82}
\begin{equation}
\label{eq:restricted-ensemble}
D^\prime_{var}=\frac{1}{Z^{\prime}_{var}}
\mbox{Tr}e^{-\beta\sum_\alpha(\varepsilon_\alpha-\mu)
a^\dagger_\alpha a_\alpha}\delta_{a^\dagger_0a_0,N_0}.
\end{equation}
with $Z^{\prime}_{var}$ chosen as usual to normalize the
distribution. In this expression we have expressed ${\hat H}_{HF}$
in a diagonal basis. This ensemble essentially excludes the {\em
dangerous} condensed mode from statistical averaging\ \cite{Iva95},
since this is what causes the problem in standard HF.  As we will
see, this  removes the spurious exchange energy among particles
occupying the condensate mode, while keeping it among particles
occupying different levels. Consistent treatment of exchange in a
Bose condensed system turns out to be essential for obtaining the
expected gapless superfluid mode of an infinite uniform system.

Using Eq.\ (\ref{eq:restricted-ensemble}), one gets the variational
free energy
\begin{align}
\Omega_{var}=-\mu N_0+\sum_{\alpha\ne0}\left[k_BT\ln(n_\alpha+1)
-\varepsilon_\alpha n_\alpha\right]+\langle\hat{H}\rangle,
\end{align}
where $\langle\hat{H}\rangle$ is
\begin{align}
\langle\hat{H}\rangle&=\sum_\alpha\varepsilon^0_{\alpha\alpha}
n_\alpha +\frac12\sum_{\alpha\beta}\bar{V}_{\alpha\beta\alpha\beta}
n_\alpha n_\beta -\frac12V_{0000}N_0^2
\delta_{\alpha,0}
\end{align}
and
\begin{align}
n_\alpha=N_0\delta_{\alpha,0}+\frac{1}{e^{-\beta(\varepsilon_\alpha-\mu)}-1}
\delta_{\alpha\ne0}.
\end{align}
When $\Omega_{var}$ is minimized with respect to $N_0$ and
$n_\alpha$ ($\alpha \ne 0$), which is equivalent to variation with
respect to $\varepsilon_\alpha$, one finds
\begin{align}
0&=-\mu+\frac{\partial\langle\hat{H}\rangle}{\partial
N_0},\ \ \alpha=0,  \\
0&=\frac{\partial\ln(n_\alpha+1)}{\partial
n_\alpha}-\varepsilon_\alpha-n_\alpha\frac{\partial\varepsilon_\alpha}
{\partial n_\alpha}+\frac{\partial\langle\hat{H}\rangle}{\partial
n_\alpha},\ \ \alpha \ne 0.
\end{align}
These equations are straightforwardly solved by setting
\begin{align}
\varepsilon_\alpha&=\varepsilon_{\alpha\alpha}^0
+\sum_\beta\bar{V}_{\alpha\beta\alpha\beta}(1-\delta_{\beta,0}
\delta_{\alpha\beta}) 
n_\beta+V_{0000}N_0\delta_{\alpha,0}
\nonumber\\
&=\varepsilon^0_{\alpha\alpha}+\Sigma_{\alpha\alpha},\ \ \text{for
all}\ \alpha
\end{align}
and
\begin{align}
\mu=\varepsilon_0,
\end{align}
where
\begin{equation}
\Sigma_{\alpha\alpha}=\sum_\beta[V_{\alpha\beta\alpha\beta}
+V_{\alpha\beta\beta\alpha}]n_\beta-V_{0000}N_0\delta_{\alpha,0}
\end{equation}
is the diagonal matrix element of the Hartree-Fock self-energy.

The problem is not fully solved because we do not have expressions
for the wavefunctions that determine the matrix elements
$V_{\alpha\beta\gamma\delta}$. To find these, we minimize the free
energy with respect to the single particle wavefunctions, keeping in
mind that they must form a complete orthonormal set\ \cite{Huse82}.
This constraint may be enforced if we write the variation in the
form
\begin{equation}
\label{eq:pert}
\delta\psi_\alpha=\sum_\gamma\eta_{\alpha\gamma}\psi_\gamma,
\text{ or } \delta
a^\dagger_\alpha=\sum_\gamma\eta_{\alpha\gamma}a^\dagger_\gamma,
\end{equation}
where $\eta_{\beta\alpha}=-\eta_{\alpha\beta}^*$. The resulting
variation of the free energy $\Omega_{var}$ may be written as
\begin{align}
\label{eq:vare-E} \delta\Omega_{var}=\sum_{\mu\ne\nu}\eta_{\nu\mu}
\langle[H,a_\mu^\dagger a_\nu]\rangle.
\end{align}
The proof of this result is given in Appendix \ref{app:delta-E}, and
is valid whether or not the system is Bose condensed. Now
substituting Eq.\ (\ref{eq:Hamiltonian-in-HF-basis}) into Eq.\
(\ref{eq:vare-E}), one gets
\begin{align}
\label{eq:E-variation}
\delta\Omega_{var}&=\sum_{\mu\ne\nu}\eta_{\nu\mu}
\left[\varepsilon^0_{\nu\mu}
+\sum_\gamma\bar{V}_{\nu\gamma\gamma\mu}n_\gamma
-V_{\nu\nu\nu\mu}n_\nu\delta_{\nu,0}\right]n_\nu
\nonumber\\
&+\sum_{\mu\ne\nu}\eta^*_{\mu\nu} \left[\varepsilon^0_{\nu\mu}
+\sum_\gamma\bar{V}_{\nu\gamma\gamma\mu}n_\gamma
-V_{\nu\mu\mu\mu}n_\mu\delta_{\mu,0}\right]n_\mu.
\end{align}
The free energy should be stationary with respect to
variations in $\eta$,
\begin{equation}
\label{eq:E-variation-0}
\frac{\delta\Omega_{var}}{\delta\eta_{\nu\mu}}
=\frac{\delta\Omega_{var}}{\delta\eta^*_{\mu\nu}}=0,
\end{equation}
which results in the equation
\begin{align}
\label{eq:sigma-off} \varepsilon^0_{\nu\mu}+
\sum_\gamma\bar{V}_{\nu\gamma\gamma\mu}n_\gamma
-N_0(V_{\nu\mu\mu\mu}\delta_{\mu,0}+V_{\nu\nu\nu\mu}\delta_{\nu,0})
=0.
\end{align}
Eq.\ (\ref{eq:sigma-off}) suggests that the self-energy matrix
should be defined as\ \cite{Huse82}
\begin{align}
\label{eq:sigma} \Sigma_{\nu\mu}&=
\sum_\gamma\bar{V}_{\nu\gamma\gamma\mu}n_\gamma
-N_0(V_{\nu\mu\mu\mu}\delta_{\mu,0}+V_{\nu\nu\nu\mu}\delta_{\nu,0})
\nonumber\\
&\ \  +N_0V_{0000}\delta_{\nu,0}\delta_{\mu,0},
\end{align}
which we note is consistent with our earlier definition
of the diagonal self-energy matrix elements.
In real space this may be written as
\begin{widetext}
\begin{align}
\label{eq:static-HF-self-energy}
\Sigma(\vec{r}_1,\vec{r}_2)&=\sum_{\nu\mu}\Sigma_{\nu\mu}
\psi_{\nu}(\vec{r}_1)\psi^*_{\mu}(\vec{r}_2)
=\delta(\vec{r}_1-\vec{r}_2)\int\!\!d\vec{r}_3V(\vec{r}_2-\vec{r}_3)
\rho(\vec{r}_3,\vec{r}_3)+V(\vec{r}_1-\vec{r}_2)\rho(\vec{r}_1,\vec{r}_2)
\nonumber\\
&\ \ -N_0\int\!\!d\vec{r}_3W_{0}(\vec{r}_3,\vec{r}_3)
\left[V(\vec{r}_1-\vec{r}_3)+V(\vec{r}_2-\vec{r}_3)\right]W_{0}
(\vec{r}_1,\vec{r}_2)+N_0V_{0000}W_0(\vec{r}_1,\vec{r}_2)
\end{align}
\end{widetext}
where $\rho(\vec{r}_1,\vec{r}_2)$ is
\begin{align}
\label{eq:rho} \rho(\vec{r}_1,\vec{r}_2)=\sum_{\alpha}n_{\alpha}
\psi_{\alpha}(\vec{r}_1)\psi^*_{\alpha}(\vec{r}_2),
\end{align}
and $W_\alpha(\vec{r}_1,\vec{r}_2)$ is defined as
\begin{align}
\label{eq:W-function}
W_{\alpha}(\vec{r}_1,\vec{r}_2)=\psi_\alpha(\vec{r}_1)
\psi^*_\alpha(\vec{r}_2).
\end{align}
With these definitions, and using the completeness of the basis,
Eq.\ (\ref{eq:sigma-off}) is equivalent to the Hartree-Fock equation
\begin{equation}
\label{eq:static-HF-eq}
\left[-\frac{\nabla^2}{2m}+U(\vec{r})\right]\psi_\alpha(\vec{r})
+\int\!\!d\vec{r}^\prime\Sigma(\vec{r},\vec{r}^\prime)
\psi_{\alpha}(\vec{r}^\prime)=\varepsilon_\alpha\psi_\alpha(\vec{r})
\end{equation}
with
\begin{equation}
\varepsilon_\alpha
=\varepsilon^0_{\alpha\alpha}+\sum_\beta\bar{V}_{\alpha\beta\beta\alpha}
n_\beta-N_0V_{0000}\delta_{\alpha,0}.
\end{equation}
Note this is consistent with Eq.\ (\ref{epsab}) expressed in a
diagonal basis.

The form of the self-energy in Eq.\
(\ref{eq:static-HF-self-energy}), as most clearly expressed in Eq.\
(\ref{eq:sigma-off}), has an interesting consequence: the
self-energy for the bosons in the condensate is different than that
of higher energy states. For the case of a uniform, infinite system
this means there is a gap between the single-particle energy of the
condensate and those of the excited states. This property of  Bose
condensates is well-known\ \cite{Leggett01}, and finding an
appropriate way to deal with it is one of the major challenges in
developing approximations for the excitation spectrum of a BEC \
\cite{Griffin96}.  For the case of a uniform infinite system, this
gap is present in the single-particle spectrum in spite of the
expected gapless collective mode spectrum.  In order to deal with
this, it is helpful to develop different Green's functions which
capture one or the other portion of the excitation spectrum, as we
now proceed to do.

\subsection{Depletion Green's Function and the TDHFA}

In formulating a TDHFA, it is useful to define Green's functions in
imaginary time and consider self-consistent approximations to their
equations of motion\ \cite{Kadanoff62}. In the fermion case, poles
of the Green's function in the absence of a time dependent potential
give the spectrum of number-changing excitations, while the response
of these Green's functions to time-dependent potentials give
collective excitations.  This allows one to conveniently separate
out these sectors of the energy spectrum. In a Bose condensed state,
these sectors become entangled in the standard Green's function
because one may add a particle to the condensate and then excite a
collective mode, yielding poles at collective mode frequencies.
Disentangling the single-particle spectrum from the collective mode
spectrum in the Green's function then becomes quite challenging.

As we now demonstrate, the TDHFA for
Bose condensates can be developed in a way that is analogous
to what works so well for fermion systems.
To to this we define a {\it depletion Green's function},
incorporating all the information about the single particle states
other than that of the condensate.
Within the static Hartree-Fock approximation this has the form
\begin{equation}
\tilde{G}({\vec r}_1,{\vec r}_2;i\omega_n) =
\sum_{\alpha \ne 0} \frac{\psi_{\alpha}({\vec r}_1)\psi_{\alpha}^*({\vec r}_2)}
{i\omega_n-{\varepsilon}_{\alpha}+\mu}.
\end{equation}
Writing this Green's function in imaginary time,
$\tilde{G}({\bf r}_1\tau_1,{\bf r}_2\tau_2)=
\frac{1}{\beta}\sum_{n}e^{-i\omega_n(\tau_1-\tau_2)}
\tilde{G}({\bf r}_1,{\bf r}_2;i\omega_n)$
one may easily show that it satisfies the equation
of motion
\begin{widetext}
\begin{align}
\left[-\frac{\partial}{\partial\tau_1} -{H}_0({\vec r}_1)
 \right]\tilde{G}({\vec r}_1\tau_1,{\vec
r}_2\tau_2)
-\int\!\! d{\vec r} ~ \Sigma({\vec r}_1,{\vec r}) \tilde{G}({\vec
r}\tau_1,{\vec r}_2\tau_2)
=\delta(\tau_1-\tau_2)[\delta({\vec r}_1-{\vec r}_2)- \psi_0({\vec
r}_1)\psi_0^*({\vec r}_2)]. \label{eq:depletionGF0}
\end{align}
By excluding the condensate state from the depletion
Green's function, we avoid the process that leads
to poles at the collective mode frequencies.  The lowest
energy poles then reflect the single-particle spectrum.

To compute collective modes of the system, it is convenient to look
at Green's functions in the presence of a time-dependent potential\
\cite{Kadanoff62}. The natural generalization of Eq.\
(\ref{eq:depletionGF0}) to this situation is
\begin{align}
\left[ -\frac{\partial}{\partial\tau_1} -{H}_0({\vec r}_1) -\delta
U({\vec r}_1,\tau_1) \right] \tilde{G}({\vec r}_1\tau_1,{\vec
r}_2\tau_2)
-\int\!\! d{\vec r} \int\!\! d\tau \Sigma({\vec r}_1 \tau_1,{\vec
r}\tau) \tilde{G}({\vec r}\tau,{\vec r}_2\tau_2)
=\delta(\tau_1-\tau_2)[\delta({\vec r}_1-{\vec r}_2)- \psi_0({\vec
r}_1,\tau_1)\psi_0^*({\vec r}_2,\tau_2)], \label{eq:depletionGF}
\end{align}
where $\delta U$ is a time-dependent potential which we will
ultimately treat perturbatively.  Note that in writing down this
equation, the self-energy Eq.\ (\ref{eq:static-HF-self-energy}) now
has time dependence, and is explicitly given by
\begin{align}
\label{eq:TDHF-self-energy}
\Sigma(\vec{r}_1\tau_1,\vec{r}_2\tau_2;\delta U)
&=\delta(\tau_1-\tau_2)\left[\delta(\vec{r}_1-\vec{r}_2)\int\!\!
d\vec{r}_3V(\vec{r}_2-\vec{r}_3)\rho(\vec{r}_3\tau_1,
\vec{r}_3\tau_1;\delta U) +V(\vec{r}_1-\vec{r}_2)
\rho(\vec{r}_1\tau_1,\vec{r}_2\tau_1;\delta U)\right]
\displaybreak[0]\nonumber\\
&-\delta(\tau_1-\tau_2)N_0\left\{\int\!\!d\vec{r}_3\left[V(\vec{r}_1-\vec{r}_3)
+V(\vec{r}_2-\vec{r}_3)\right]
W_{0}(\vec{r}_3\tau_1,\vec{r}_3\tau_1;\delta U)-V_{0000}\right\}
W_{0}(\vec{r}_1\tau_1,\vec{r}_2\tau_1;\delta U).
\end{align}
The time dependence enters through the {\it wavefunctions} in the
quantities $\rho$ and $W_{\alpha}$ (Eqs.\ (\ref{eq:rho}) and
(\ref{eq:W-function})), and we have noted that these quantities are
now functionals of $\delta U$ which is ultimately responsible for
the time dependence.

It is useful to note at this point that we have made a crucial
assumption, which can be understood as the essential underlying
approximation of the TDHFA: we allow only the wavefunctions to
change with time, while the occupations $n_{\alpha}$ remain
stationary and equal to their values for $\delta U=0$.  This can be
shown\ \cite{Blaizot86} to be equivalent to an assumption that the
entropy of the system remains unchanged in the presence of $\delta
U$.

Noting that Eq. \ref{eq:rho} may be recast in the form
\begin{equation}
\rho({{\vec r}_1\tau_1,\vec r}_2\tau_2)= \left[
\tilde{G}(\vec{r}_2,\tau_1^+,\vec{r}_1,\tau_1)
+N_0W_0({\vec r}_1\tau_1,{\vec r}_2\tau_1) \right] \delta(\tau_1-\tau_2),
\label{eq:rhot}
\end{equation}
with $W_0({\vec r}_1\tau_1,{\vec r}_2\tau_1)= \psi_0({\vec
r}_1\tau_1)\psi_0^*({\vec r}_2\tau_2)$, we see that Eqs.\
(\ref{eq:static-HF-self-energy}), (\ref{eq:depletionGF}) and
(\ref{eq:rhot}) nearly form a closed set of equations.  We have left
to determine the time dependence of $\psi_0$.  An important aspect
of the problem is to assure that our TDHFA obeys particle
conservation, a feature that is often difficult to build into
collective mode calculations for Bose condensates\ \cite{Griffin96}.
In the present case we can guarantee this by having the condensate
wavefunction controlled by the same effective Hamiltonian as the
excited states, via Eq.\ (\ref{eq:depletionGF}).  Thus we take
\begin{equation}
\left[ -\frac{\partial}{\partial\tau_1} -{H}_0({\vec r}_1)
-\delta U({\vec r}_1,\tau_1) \right] \psi_0({\vec r}_1\tau_1)
-\int\!\! d{\vec r} \int\!\! d\tau \Sigma({\vec r}_1 \tau_1,{\vec
r},\tau) \psi_0({\vec r}\tau) =0. \label{wf0}
\end{equation}
\end{widetext}
With this equation for $\psi_0$, it is easy to verify that the
overlaps $\int d{\bf r} \psi_0^*({\bf r} \tau) \tilde{G} ({\bf r}
\tau,{\bf r}^{\prime} \tau^{\prime})$ and $\int d{\bf r}^{\prime}
\tilde{G} ({\bf r} \tau,{\bf r}^{\prime} \tau^{\prime}) \psi_0({\bf
r}^{\prime} \tau^{\prime})$ vanish, so
that the depletion Green's function involves no
change in the number of condensate particles even in
the presence of the time-dependent potential.
It is the possibility of changing this and simultaneously
creating a collective excitation that allows the
collective mode spectrum to appear in the standard Green's function.
Thus by working with the
depletion Green's function we avoid the entangling of
particle-conserving excitations and single-particle excited states
that characterize the approaches based on broken gauge symmetry.

In principle these equations may be solved self-consistently to
develop a mean-field approximation for this many-body system in a
time-dependent potential.  In practice this is a formidable task, so
one instead focuses on the linear response of $\tilde{G}$ and
$\psi_0$ to small perturbations $\delta U$.  These may be used to
construct, for example, the density response function, whose poles
give the collective modes of the system\ \cite{Kadanoff62}.

\begin{widetext}
\section{Linear Response}
\label{sec:chi}

\subsection{General Formulation}
We begin by expanding Eq. \ref{eq:depletionGF} for small $\delta U$,
retaining only terms that are linear in this quantity.  This
leads to the equation
\begin{align}
\label{eq:deltaG}
\delta U({\vec r}_1\tau_1)&\overline{W}_0({\vec r}_1,{\vec r}_2)
\delta(\tau_1-\tau_2) =
\left[\frac{\partial}{\partial \tau_2} - \hat{H}_{HF}({\vec
r}_2)\right]\delta(\tau_1-\tau_2)\delta W_0({\vec r}_1,{\vec
r}_2;\tau_1)
\nonumber\\
&+\left[-\frac{\partial}{\partial \tau_1} - \hat{H}_{HF}({\vec
r}_1)\right]\left[\frac{\partial}{\partial \tau_2} -
\hat{H}_{HF}({\vec r}_2)\right] \delta\tilde{G}({\vec r}_1,\tau_1;{\vec
r}_2,\tau_2)
-\int\!\! d{\vec r}_3 \delta\Sigma({\vec r}_1,\tau_1;{\vec r}_2,\tau_2)
\overline{W}_0({\vec r}_3,{\vec r}_2)\delta(\tau_1-\tau_2),
\end{align}
where $\overline{W}_0({\vec r}_1,{\vec r}_2) =
\delta({\vec r}_1-{\vec r}_2) - W_0({\vec r}_1,{\vec r}_2)$,
and $\delta W_0({\vec r}_1,{\vec r}_2;\tau)
=\delta\psi_0({\vec r}_1,\tau)\psi_0^*({\vec r}_2)+
\psi_0({\vec r}_1)\delta\psi_0^*({\vec r}_2,\tau)$.  The operator
$\hat{H}_{HF}({\vec r})$ has the meaning, for example,
$$\hat{H}_{HF}({\vec r}_2) \delta W_0({\vec r}_1,{\vec r}_2;\tau_1) =
H_0({\vec r}_2) \delta W_0({\vec r}_1,{\vec r}_2;\tau_1) +\int\!\!
d{\vec r} \Sigma({\vec r}_2,{\vec r}) \delta W_0({\vec r}_1,{\vec
r};\tau_1)$$
\end{widetext}
with $\Sigma$ given by Eq.\ (\ref{eq:static-HF-self-energy}). The
variation of the self-energy, $\delta\Sigma$, comes from the fact
that the wavefunctions $\psi_{\alpha}$ are functionals of $\delta
U$; we will provide an explicit expression for the specific case of
a contact potential below.  Before proceeding with this, we
demonstrate that the ideas developed above may be used to compute an
important quantity, the density response function.

\subsection{Density Response Function}

An alternative procedure for avoiding the singularity
in the standard imaginary time Green's function is to
work directly with the wavefunctions and the
density matrix
$$\rho(\vec{r}_1,\vec{r}_2,\tau;\delta U)=
\sum_{\alpha}n_{\alpha}
\psi_{\alpha}(\vec{r}_1,\tau)\psi^*_{\alpha}(\vec{r}_2,\tau),
$$
which is a generalization of Eq.\ (\ref{eq:rho}). In this context it
is convenient to work with real rather than imaginary time. The
equation of motion for real time density matrix
$\rho(\vec{r}_1,\vec{r}_2,t;\delta U)$ can be easily obtained from
the time-dependent Hartree-Fock equations for the wavefunctions,
\begin{widetext}
\begin{align}
\label{eq:HF-eq-T-finite} i\frac{\partial}{\partial
t_1}\psi_\alpha(\vec{r}_1t_1)&=\left[-\frac{\nabla^2_1}{2m}
+U(\vec{r}_1)+\delta U(\vec{r}_1t_1)\right]\psi_\alpha(\vec{r}_1t_1)
+\int\!\!d\vec{r}_2dt_2\Sigma(\vec{r}_1t_1,\vec{r}_2 t_2;\delta
U)\psi_\alpha(\vec{r}_2 t_2),
\end{align}
with the result
\begin{align}
\label{eq:G-eq-3} 0&=\left[i\frac{\partial}{\partial t}
+\left(\frac{\nabla^2_1}{2m}-\frac{\nabla^2_2}{2m}\right)
-U(\vec{r}_1t)+U(\vec{r}_2t)\right]
\rho(\vec{r}_1,\vec{r}_2,t;\delta U) \nonumber\\&\ \ \
-\int\!\!d\vec{r}_3 \left[\Sigma(\vec{r}_1t,\vec{r}_3t;\delta U)
\rho(\vec{r}_3,\vec{r}_2,t;\delta U)
-\rho(\vec{r}_1,\vec{r}_3,t;\delta U)
\Sigma(\vec{r}_3t,\vec{r}_2t;\delta U)\right].
\end{align}
\end{widetext}
In writing this equation we have analytically continued the
self-energy to real time, which in practice simply involves
replacing $\tau \rightarrow t$ in all the arguments of Eq.\
(\ref{eq:TDHF-self-energy}).

Expanding
$\rho(\vec{r}_1,\vec{r}_2,t;\delta U)$,
$\Sigma(\vec{r}_1t,\vec{r}_2t;\delta U)$ and
$W_0(\vec{r}_1t,\vec{r}_2t;\delta U)$ around their static HF values
and retaining only terms that are
first order in $\delta U(\vec{r}t)$, one obtains a
linearized equation for $\delta\rho$,
\begin{widetext}
\begin{align}
\label{eq:linearized-G} \left[\delta U(\vec{r}_1t)-\delta
U(\vec{r}_2t)\right]\rho(\vec{r}_1,\vec{r}_2)&=
\left[i\frac{\partial}{\partial
t}-\hat{H}_{HF}(\vec{r}_1)+\hat{H}_{HF}(\vec{r}_2)\right]
\delta\rho(\vec{r}_1t,\vec{r}_2t^+;U)
\nonumber\\
&\ \ \
-\int\!\!d\vec{r}_3\left[\delta\Sigma(\vec{r}_1t,\vec{r}_3t;\delta
U) \rho(\vec{r}_3,\vec{r}_2)-\rho(\vec{r}_1,\vec{r}_3)
\delta\Sigma(\vec{r}_3t,\vec{r}_2t;\delta U)\right].
\end{align}
A density matrix response function\ \cite{Kadanoff62} may be defined
as
\begin{align}
\chi^R(\vec{r}_1\vec{r}_3,\vec{r}_2\vec{r}_3,t-t_3)
&=\theta(t-t_3)\left.\frac{\delta\rho(\vec{r}_1,\vec{r}_2,t;\delta
U)}{\delta U(\vec{r}_3t_3)}\right|_{\delta U=0},
\end{align}
with the more standard density-density response function then given
by $\chi^R(\vec{r}_1\vec{r}_3,\vec{r}_1\vec{r}_3,t-t_3)$. One can
rewrite Eq.\ (\ref{eq:linearized-G}) as
\begin{align}
\label{eq:chi} \left[i\frac{\partial}{\partial t}
-\hat{H}_{HF}(\vec{r}_1)+\hat{H}_{HF}(\vec{r}_2)\right]
&\chi^R(\vec{r}_1\vec{r}_4,\vec{r}_2\vec{r}_4,t-t_4)=
\left[\delta(\vec{r}_1-\vec{r}_4)-\delta(\vec{r}_2-\vec{r}_4)\right]
\delta(t-t_4)\rho(\vec{r}_1,\vec{r}_2)
\nonumber\\
&+\int\!\!d\vec{r}_3\left[\Gamma^R(\vec{r}_1\vec{r}_4,\vec{r}_3\vec{r}_4,t-t_4)
\rho(\vec{r}_3,\vec{r}_2)-\rho(\vec{r}_1,\vec{r}_3)\Gamma^R(\vec{r}_3\vec{r}_4,
\vec{r}_2,\vec{r}_4,t-t_4)\right],
\end{align}
where
\begin{align}
\label{eq:gamma-function}
\Gamma^R(\vec{r}_1\vec{r}_4,\vec{r}_2\vec{r}_4,t-t_4)
&=\theta(t-t_4)\left.\frac{\delta\Sigma(\vec{r}_1t,\vec{r}_2t,\delta
U)}{\delta U(\vec{r}_4t_4)}\right|_{\delta U=0}
\nonumber\\
&=\delta(\vec{r}_1-\vec{r}_2)\int\!\!d\vec{r}_3V(\vec{r}_2-\vec{r}_3)
\chi^R(\vec{r}_3\vec{r}_4,\vec{r}_3\vec{r}_4,t-t_4)
+V(\vec{r}_1-\vec{r}_2)\chi^R(\vec{r}_1\vec{r}_4,\vec{r}_2\vec{r}_4,t-t_4)
\displaybreak[0]\nonumber\\
&-N_0\int\!\!d\vec{r}_3\left[V(\vec{r}_1-\vec{r}_3)+V(\vec{r}_2-\vec{r}_3)\right]
\left.\frac{\delta W_{0}(\vec{r}_3t,\vec{r}_3t;\delta U)}{\delta
U(\vec{r}_4t_4)}\right|_{\delta U=0}W_{0}(\vec{r}_1,\vec{r}_2)
\displaybreak[0]\nonumber\\
&-N_0\left\{\int\!\!d\vec{r}_3\left[V(\vec{r}_1-\vec{r}_3)+V(\vec{r}_2-\vec{r}_3)\right]
W_{0}(\vec{r}_3,\vec{r}_3)-V_{0000}\right\}\left.\frac{\delta
W_{0}(\vec{r}_1t,\vec{r}_2t;\delta U)}{\delta
U(\vec{r}_4t_4)}\right|_{\delta U=0}
\displaybreak[0]\nonumber\\
&+2N_0\int\!\!d\vec{r}_3\int\!\!d\vec{r}_5V(\vec{r}_3-\vec{r}_5)
\left.\frac{\delta W_{0}(\vec{r}_3t,\vec{r}_3t;\delta U)}{\delta
U(\vec{r}_4t_4)}\right|_{\delta U=0}W_{0}(\vec{r}_5,\vec{r}_5)
W_{0}(\vec{r}_1,\vec{r}_2).
\end{align}
The functional derivative of $W_0$ with respect to $\delta
U(\vec{r}t)$ requires one to work directly with the wavefunctions.
This is discussed in detail in Appendix B, with the result (Eq.\
(\ref{eq:phi-perturbation}))
\begin{align}
\left.\label{eq:W-U} \frac{\delta W_0(\vec{r}_1t_1,\vec{r}_1^\prime
t_1;\delta U)} {\delta U(\vec{r}_2t_2)}\right|_{\delta U=0}
&=-\sum_{\beta\lambda\mu}\int\!\!\frac{d\omega}{2\pi}
\frac{e^{-i\omega(t_1-t_2)}}
{\omega+\varepsilon_{\beta}-\varepsilon_{0}}
\left[\delta_{0\lambda}\delta_{\beta\mu}
+\Gamma^R_{0\beta,\lambda\mu}(\omega)\right]
\psi^*_\lambda(\vec{r}_2)\psi_\mu(\vec{r}_2)
\psi_0(\vec{r}_1)\psi^*_\beta(\vec{r}_1^\prime)
\nonumber\\
&\ \ \ +\sum_{\beta\lambda\mu}\int\!\!\frac{d\omega}{2\pi}
\frac{e^{-i\omega(t_1-t_2)}}
{\omega+\varepsilon_{0}-\varepsilon_{\beta}} \left[
\delta_{0\mu}\delta_{\beta\lambda} +\Gamma^R_{\beta0,\lambda
\mu}(\omega)\right] \psi_\lambda^*(\vec{r}_2)\psi_\mu(\vec{r}_2)
\psi_\beta(\vec{r}_1)\psi^*_0(\vec{r}_1^\prime).
\end{align}

To make further progress, it is convenient to expand
the functions $\chi^R$ and $\Gamma^R$ in terms of
the static HF eigenstates $\psi_{\alpha}$ as
\begin{align}
\chi^R(\vec{r}_1\vec{r}_3,\vec{r}_2\vec{r}_3,t-t_3)
&=\sum_{\substack{\alpha\beta\\
\lambda\mu}}\int_{-\infty}^{\infty}\frac{d\omega}{2\pi}
\chi_{\alpha\beta,\lambda\mu}(\omega)e^{-i\omega(t-t_3)}\psi^*_{\beta}(\vec{r}_2)
\psi_\alpha(\vec{r}_1)\psi^*_\mu(\vec{r}_3)\psi_\lambda(\vec{r}_3),
\nonumber\\
\Gamma^R(\vec{r}_1\vec{r}_3;\vec{r}_2\vec{r}_3,t_1-t_3)
&=\sum_{\substack{\alpha\beta\\
\lambda\mu}}\int_{-\infty}^{\infty}\frac{d\omega}{2\pi}
\psi^*_\alpha(\vec{r}_2)\psi_\beta(\vec{r}_1)
\psi_{\lambda}^*(\vec{r}_3)\psi_{\mu}(\vec{r}_3)
\Gamma^R_{\beta\alpha,\lambda\mu}(\omega)e^{-i\omega(t_1-t_3)}.
\end{align}

Substituting these expansions into Eqs.\ (\ref{eq:chi}),
(\ref{eq:gamma-function}), and (\ref{eq:W-U}), one obtains
\begin{align}
\label{eq:chi-eigen}
\left(\omega+\varepsilon_\mu-\varepsilon_\lambda\right)
\chi^R_{\lambda\mu,\alpha\beta}(\omega)=\left(n_\mu-n_\lambda\right)
\left[\delta_{\lambda\alpha}\delta_{\mu\beta}
+\Gamma^R_{\lambda\mu,\alpha\beta}(\omega)\right],
\end{align}
and
\begin{align}
\label{eq:gamma-eigen}
\Gamma_{\lambda\mu,\alpha\beta}(\omega)&=\sum_{\lambda_1\mu_1}
\left[V_{\lambda\mu_1\lambda_1\mu}+V_{\lambda\mu_1\mu\lambda_1}\right]
\chi^R_{\lambda_1\mu_1,\alpha\beta}(\omega) \nonumber\\
&-\sum_{\lambda_1\mu_1}\delta_{\lambda_10}\left[V_{\lambda\mu_1\mu\lambda_1}
\delta_{\lambda_1\mu}+V_{\lambda\mu_1\lambda_1\mu}\delta_{\lambda_1\lambda}\right]
\chi^R_{\lambda_1\mu_1,\alpha\beta}(\omega) \nonumber\\
&-\sum_{\lambda_1\mu_1}\delta_{\mu_10}\left[V_{\lambda\mu_1\lambda_1\mu}\delta_{\mu_1\mu}
+V_{\lambda\mu_1\mu\lambda_1}\delta_{\mu_1\lambda}\right]
\chi^R_{\lambda_1\mu_1,\alpha\beta}(\omega)
\nonumber\\
&-\sum_{\lambda_1\mu_1}
\left(\delta_{\lambda_10}+\delta_{\mu_10}\right)
\left(\delta_{\lambda_1\lambda}V_{\mu_100\mu}+\delta_{\mu_1\mu}V_{\lambda00\lambda_1}\right)
\chi^R_{\lambda_1\mu_1,\alpha\beta}(\omega)
\nonumber\\
&+N_0V_{0000}\left(\delta_{\lambda0}-\delta_{\mu0}\right)\chi^R_{\lambda\mu,\alpha\beta}
(\omega).
\end{align}
Using Eq.\ (\ref{eq:gamma-eigen}) in Eq.\ (\ref{eq:chi-eigen}) to
eliminate $\Gamma$, one ultimately may write what amounts to a
matrix equation for $\chi^R_{\lambda\mu,\alpha\beta}(\omega)$,
\begin{align}
\label{eq:chi-eq} (\omega+\varepsilon_\mu
-\varepsilon_\lambda)&\chi^R_{\lambda\mu,\alpha\beta}(\omega)
=(n_\mu-n_\lambda)
\delta_{\alpha\lambda}\delta_{\beta\mu}+(n_\mu-n_\lambda)
\sum_{\lambda_1\mu_1}\left[V_{\lambda\mu_1\lambda_1\mu}
+V_{\lambda\mu_1\mu\lambda_1}\right]\chi^R_{\lambda_1\mu_1,\alpha\beta}(\omega)
\nonumber\\
&-(n_\mu-n_\lambda)
\sum_{\lambda_1\mu_1}\left[V_{\lambda\mu_1\lambda_1\mu}
\left(\delta_{\lambda_1,0}\delta_{\lambda,\lambda_1}
+\delta_{\mu_1,0}\delta_{\mu,\mu_1}\right)
+V_{\lambda\mu_1\mu\lambda_1}\left( \delta_{\lambda,0}
+\delta_{\mu,0}\right)\left( \delta_{\lambda_1,\lambda}
+\delta_{\mu_1,\mu} \right)\right]
\chi^R_{\lambda_1\mu_1,\alpha\beta}(\omega).
\end{align}
\end{widetext}
Interestingly, the first line this equation has the same form as
what is obtained from the TDHFA using the standard grand canonical
ensemble, although the energies and occupations are different
because of the non-local terms in the self-energy that arise from
using the constrained grand-canonical ensemble. The terms in the
second line also come from these non-local terms. This equation for
the density response function provides a natural way to describe the
coupling between the condensate and normal components through the
matrix indices, which may refer to the condensate state or to the
depletion states. We will illustrate this explicitly when we solve
Eq.\ (\ref{eq:chi-eq}) for a homogenous system below.

Finally, it is useful to note that Eq.\ (\ref{eq:chi-eq}) can
written in the form of a Bethe-Salpeter equation
\begin{align}
\chi_{\lambda\mu,\alpha\beta}(\omega)&=\chi^0_{\lambda\mu,\alpha\beta}(\omega)
\nonumber\\&
+\sum_{\eta\nu\rho\sigma}\chi^0_{\lambda\mu,\rho\nu}(\omega)
K_{\rho\nu,\eta\sigma}(\omega)\chi_{\eta\sigma,\alpha\beta}(\omega),
\end{align}
where $\chi^0_{\lambda\mu,\alpha\beta}(\omega)$ is defined as
\begin{align}
\label{eq:chi0-eq}
\chi^0_{\lambda\mu,\alpha\beta}(\omega)=\frac{n_{\mu}-n_{\lambda}}
{\omega+i\eta+\varepsilon_\mu-\varepsilon_\lambda}\delta_{\alpha\lambda}
\delta_{\beta\mu},
\end{align}
and the kernel $K$ as
\begin{align}
K_{\rho\nu,\eta\sigma}(\omega)&=V_{\rho\sigma\eta\nu}\left(1-
\delta_{\eta0}\delta_{\rho\eta}-\delta_{\sigma0}\delta_{\sigma\nu}\right)
\nonumber\\
&+V_{\rho\sigma\nu\eta}\left[1-\left(\delta_{\rho0}+\delta_{\nu0}\right)
\left(\delta_{\eta\rho}+\delta_{\sigma\nu}\right)\right]
\end{align}
which is independent of the frequency.

Equation (\ref{eq:chi-eq}) is valid for a dilute Bose gas with any
shape of static external potential $U(\vec{r})$ and a general
two-body interaction potential $V(\vec{r}_1-\vec{r}_2)$.  In general
it must be solved numerically. In the next section, we will show
that it can be solved analytically for a uniform, homogeneous system
of bosons interacting via a contact two-body interaction.

\section{Application to a Homogenous System}
\label{sec:homogenous}

In this section, we apply the above results to discuss the
single-particle and collective excitations of a homogeneous system
with a contact two-body interaction, $V(\vec{r}_1-\vec{r}_2)
=g\delta(\vec{r}_1-\vec{r}_2)$.  This is a common and quantitatively
accurate approximation for dilute Bose gases in many situations.  We
begin by reviewing the solutions of the static Hartree-Fock
equations for this case\ \cite{Pethick02}.

\subsection{Static Hartree-Fock for a Bose-Einstein Condensate}

For a uniform homogeneous system ($U({\vec r})=0$), the
Hartree-Fock Hamiltonian is diagonalized by plane waves,
$\psi_{\vec k}({\vec r})=\frac{1}{\sqrt{\Omega}}e^{i {\vec k} \cdot {\vec r}}$,
with $\Omega$ the volume of the system,
and the single-particle energies are easily shown to be
\begin{equation}
\varepsilon_{\vec{k}}
=\varepsilon^0_{\vec{k}}+2g\rho-g\rho_0\delta_{\vec{k},0}
\end{equation}
where $\varepsilon^0_{\vec{k}}=\frac{\hbar^2\vec{k}^2}{2m}$,
$\rho=N/\Omega$ and $\rho_0=N_0/\Omega$.
Below the critical temperature, the chemical potential is given
by the lowest eigenvalue
\begin{equation}
\mu=\varepsilon_0=g(2\rho-\rho_0).
\end{equation}
Note that there is a gap in the single-particle spectrum
$\Delta=g\rho_0$ with respect to the chemical potential for a Bose condensed
system. As discussed in the Introduction, if the grand canonical ensemble
is used, there is no such gap and the corresponding single-particle energy
is $\varepsilon_{\vec{k}} =\varepsilon^0_{\vec{k}}+2g\rho$.
However the absence of the gap is an artifact of strong number fluctuations
in the condensed state, which are more properly controlled by
the constrained grand canonical ensemble.

The gapped structure of the single-particle spectrum may be probed
in principle via a tunneling experiment.  One interesting possible
approach involves a ``single-atom pipette'' that has recently been
proposed\ \cite{kolomeisky}, in which bosons are loaded into a
strongly localized potential. Because of the potential well, there
can be a considerable gap for excitations within the pipette, and at
low temperatures all the atoms will be at the same energy. By
bringing the pipette sufficiently close to the bulk BEC for a fixed
time, atoms may tunnel out of the pipette into the bulk BEC, and a
measurement of the number of atoms remaining in the pipette after
this process allows one to infer the tunneling rate.  For a
relatively deep pipette potential, one expects the atoms to tunnel
only into the condensate state of the BEC. As the energy is raised,
however, there should be a sharp threshold energy, set by the gap,
above which atoms may also tunnel into the excited states. This will
show up as a non-analytic contribution to the tunneling rate
reflecting the density of depletion states in the BEC.  For a bulk,
three-dimensional BEC, this would be proportional to
$(E-E_{th})^{1/2}$, with $E$ the energy of the bosons in the
pipette, and $E_{th}$ set by the gap energy. We note that the
exponent might be renormalized by shakeup effects\ \cite{Mahan}, but
we expect the non-analytic behavior to be robust and to reflect the
gap in the single-article spectrum.

The critical temperature temperature $T_c$ is determined by taking
the limit $\rho_0 \rightarrow 0$, so that the number of particles
depleted from the condensate tends to the total number of particles $N$.
This leads to the condition
\begin{equation}
N=\frac{V}{(2\pi)^3}\int d^3\vec{k}
\frac{1}{e^{\beta_c\varepsilon^0_{\vec{k}}}-1},
\end{equation}
which gives \cite{Pethick02}
\begin{align}
k_BT_c=\frac{4\pi}{2m}\rho^{2/3}\left(\frac{1}{g_{3/2}(1)}
\right)^{2/3},
\end{align}
where
\begin{align}
 g_{\gamma}(x)&=\sum_{n=1}^{\infty}\frac{x^n}{n^\gamma}.
\end{align}

\begin{figure}[h]
\includegraphics[width=1\columnwidth]{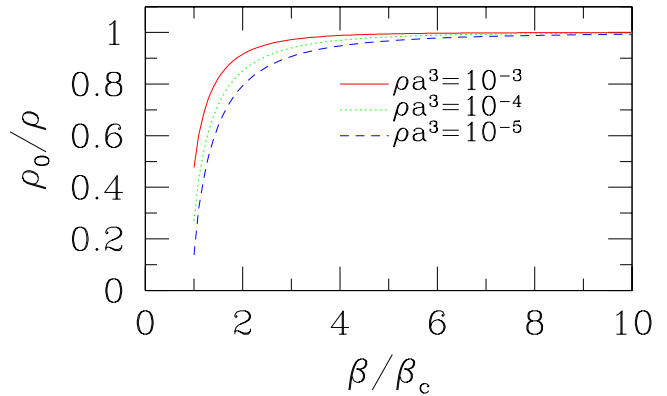}
\vspace{-5mm}\caption{(Color online) Condensation fraction as a
function of the temperature at two dilute parameters. }
\label{fig:n0}
\end{figure}

The condensate density $\rho_0$ below $T_c$ is found by requiring
depletion of the condensate to be equal to the thermal occupation of
the excited states\ \cite{Pethick02},
\begin{align}
\label{eq:rho0-T}
1-\frac{\rho_0}{\rho}&=\frac{1}{(2\pi)^3\rho}\int\!\!d\vec{k}\frac{1}
{e^{\beta(\varepsilon^0_{\vec{k}}+g\rho_0)}-1}
=\frac{1}{\lambda_T^{\prime3}}g_{3/2}(z),
\end{align}
where $z=e^{-\beta g\rho_0}$, and
\begin{align}
\lambda^\prime_T& =\left(\frac{2\pi\hbar^2\beta}{m}\right)^{1/2}
\rho^{1/3}=\lambda_T\rho^{1/3}
\end{align}
is the product of the thermal wavelength and the inverse average
distance between particles.  We note that although the Hartree-Fock
approximation yields precisely the same $T_c$ as for a
non-interacting system, the phase transition into the condensed
state is first-order rather than continuous. This can be seen in the
behavior of the condensation fraction as a function of the
temperature, as illustrated in Fig.\ \ref{fig:n0}.

From Eq.\ (\ref{eq:rho0-T}) and Fig.\ \ref{fig:n0}, we can see that
in the HFA, the repulsive interaction enhances the condensation
relative to the non-interacting case\ \cite{Leggett01}.

\subsection{Collective Modes from the Depletion Green's Function}

We would like to solve Eq.\ (\ref{eq:deltaG}) for the case at hand;
in order to do this we need to specify $\delta\Sigma$.  For contact
interactions in the homogeneous system, this is easily evaluated,
with the result
\begin{widetext}
\begin{equation}
\delta\Sigma({\vec r}_1,\tau_1;{\vec r}_2,\tau_2)= \left\lbrace
2g\left[\delta\tilde{\rho}({\vec r_1},\tau_1)+ \delta\rho_0({\vec
r}_1,\tau_1)\right]\delta({\vec r}_1-{\vec r}_2)
+g\left[\delta w({\vec r}_1,{\vec
r}_2;\tau_1)\rho_0({\vec r}_1,{\vec r}_2)
-\frac{1}{\Omega}\delta \rho_0({\vec r}_1,{\vec
r}_2;\tau_1)\right]\right\rbrace\delta(\tau_1-\tau_2).
\label{ds_uniform}
\end{equation}
In this equation, $\delta\tilde{\rho}({\vec r_1},\tau_1)=
\delta\tilde{G}({\vec r}_1\tau_1,{\vec r}_1\tau_1^+)$
is the variation of the depleted particle density,
$\delta\tilde{\rho}_0({\vec r_1},\tau_1)=N_0\delta W_0({\vec r}_1,{\vec r}_1,\tau_1)$
is the variation of the condensate density, and
$\delta w({\vec r}_1,{\vec
r}_2;\tau_1)=-[\delta \psi_0(\vec{r}_1)+\delta \psi_0^*(\vec{r}_1)
+\psi_0(\vec{r}_2)+\delta \psi_0^*(\vec{r}_2)]/\sqrt{\Omega}$.
As in the case of the density response function, we can make progress
by expanding the various quantities in terms of the unperturbed
single particle states.  Thus we define
$\delta\tilde{G}_{{\vec k}_1{\vec k}_2}(i\omega_1,i\omega_2)$
via
\begin{equation}
\delta\tilde{G}({\vec r}_1\tau_1,\vec{r}_2\tau_2)=\frac{1}{\beta^2}
\sum_{i\omega_1,i\omega_2}e^{-i\omega_1\tau_1+i\omega_2\tau_2}
\frac{1}{\Omega}\sum_{\vec{k}_1,\vec{k}_2}e^{i\vec{k}_1\cdot \vec{r}_1
-i\vec{k}_2\cdot \vec{r}_2}
\delta\tilde{G}_{\vec{k}_1,\vec{k}_2}(i\omega_1,i\omega_2).
\label{ft}
\end{equation}
It is useful to notice that the Fourier transform of the
depletion density, $\delta\tilde{\rho}({\vec k},i\omega_n)=
\int_0^\beta d\tau \int d\vec{r} e^{i\vec{k}\cdot {\vec r}+
i\omega_n\tau} \delta\tilde{\rho}(\vec{r},\tau)$, satisfies
\begin{equation}
\delta\tilde{\rho}({\vec k}_,i\omega_n)=
\frac{1}{\Omega\beta}\sum_{i\omega_n',{\vec k}'}
\delta\tilde{G}_{{\vec k}+{\vec k}',{\vec k}'}(i\omega_n+i\omega_n',i\omega_n').
\label{depletion_density}
\end{equation}
Using Eq.\ (\ref{ft}), Eq.\ (\ref{eq:deltaG}) may be recast in a
particularly simple form.  For $\vec{k}_1,\vec{k}_2 \ne 0$, we find
\begin{equation}
[\varepsilon_{\vec{k}_1}-i\omega_1][\varepsilon_{\vec{k}_2}-i\omega_2]
\delta\tilde{G}_{\vec{k}_1,\vec{k}_2}(i\omega_1,i\omega_2)
+\frac{2g}{\Omega\beta} \sum_{\vec{k}',i\omega'}
\delta\tilde{G}_{\vec{k}'+\Delta\vec{k},\vec{k}'}
(i\omega'+i\Delta\omega,i\omega')
+2g\delta\rho_0(\Delta\vec{k},i\Delta\omega)
=-\delta U(\Delta\vec{k},i\Delta\omega),
\label{den1}
\end{equation}
where $\Delta\omega=\omega_1-\omega_2$, $\Delta\vec{k}=\vec{k}_1-\vec{k}_2$,
and
\begin{align}
\delta\rho_0(\Delta\vec{k},i\Delta\omega_n)=&\frac{1}{\Omega}
\int_0^{\beta}d\tau \int d\vec{r}
e^{i\Delta\vec{k}\cdot\vec{r}+i\Delta\omega_n\tau}
[\delta\psi_0(\vec{r},\tau)
+\delta\psi_0^*(\vec{r},\tau)]
\nonumber\\
\delta U(\Delta\vec{k},i\Delta\omega_n)=&\frac{1}{\Omega}
\int_0^{\beta}d\tau \int d\vec{r}
e^{i\Delta\vec{k}\cdot\vec{r}+i\Delta\omega_n\tau}
\delta U({\vec r},\tau).
\nonumber
\end{align}
It is clear that in Eq.\ (\ref{den1}) we would like to write
$\vec{k}_1=\vec{k}_2+\Delta\vec{k}$,
$\omega_1=\Delta\omega+\omega_2$, divide by the equation by
$[\varepsilon_{\vec{k}_1}-i\omega_1][\varepsilon_{\vec{k}_2}-i\omega_2]$
and sum with respect to $\vec{k}_1$ and $\omega_1$.  However, there
is a caveat: in doing this our indices run over $k_1$=0 and $k_2=0$,
which introduces further terms in Eq.\ (\ref{den1}).  It is not
difficult to show that these terms vanish in the thermodynamic limit
($\Omega \rightarrow \infty$), which is the case of interest to us.
Performing the above steps, Eq.\ (\ref{den1}) becomes a simple
linear equation,
\begin{equation}
\left[ \frac{1}{\tilde{P}(\Delta\vec{k},i\Delta\omega)}-2g\right]
\delta \tilde{\rho}(\Delta\vec{k},i\Delta\omega)
-2g\delta \rho_0 (\Delta\vec{k},i\Delta\omega) = \delta U(\Delta\vec{k},i\Delta\omega),
\label{first_eq}
\end{equation}
where
$\tilde{P}(\vec{q},i\omega_n)=\frac{1}{\Omega}\sum^{\prime}_{\vec k}
[n_B(\tilde{\varepsilon}_{{\vec k}+{\vec
q}})-n_B(\tilde{\varepsilon}_{{\vec k}})]/
[i\omega_n+\tilde{\varepsilon}_{{\vec k}+{\vec
q}}-\tilde{\varepsilon}_{{\vec k}}]$,
$\tilde{\varepsilon}_{{\vec k}}=\varepsilon_{{\vec k}}-\mu$,
$n_B(\varepsilon)=1/(e^{\beta\varepsilon}-1)$ is the Bose occupation
factor, and the prime on the sum indicates that ${\vec k},{\vec
k}+{\vec q}=0$ should not be included.

Since the problem involves two density disturbances,
the depletion density $\delta\tilde{\rho}$ and the
condensate density $\delta\rho_0$, we need a second
equation.  This can be obtained directly from the
groundstate wavefunction disturbance, as described
in Appendix B, since $\delta\rho_0({\vec r},\tau)=
\frac{N_0}{\sqrt{\Omega}}\left[ \delta\psi_0({\vec r},\tau)+
\delta\psi_0^*({\vec r},\tau) \right]$.  Using the method
of Appendix B, one may easily show for imaginary time,
\begin{equation}
\delta\psi_0({\vec r},i\omega_n)=
\frac{1}{\Omega^{3/2}}{\sum_{\vec{k}}}^\prime \left[ \frac{\delta
U(\vec{k},i\omega_n) + \delta\Sigma(\vec{k},i\omega_n)} {i\omega_n -
\tilde{\varepsilon}_{\vec k}} \right]e^{i\vec{k}\cdot \vec{r}},
\label{dpsi0}
\end{equation}
where the prime on the sum indicates $\vec{k}=0$ should not be included, and
\begin{equation}
\delta\Sigma(\vec{k},i\omega_n)=
\int d\vec{r}_1 d\vec{r}_2 \int d\tau_1 d\tau_2
e^{i\omega_n\tau_1+i\vec{k}\cdot \vec{r}_1}
\delta\Sigma(\vec{r}_1\tau_1,\vec{r}_2\tau_2).
\end{equation}
Substituting Eq.\ (\ref{ds_uniform}) into Eq.\ (\ref{dpsi0}), one
obtains
\begin{equation}
\delta\psi_0(\vec{k},i\omega_n)=
\frac{1}{i\omega_n-\tilde{\varepsilon}_{\vec k}}
\left[
\frac{1}{\sqrt{\Omega}}\delta U(\vec{k},i\omega_n)
+ \frac{2g}{\sqrt{\Omega}} \delta \tilde{\rho}(\vec{k},i\omega_n)
+ g\rho_0 \delta\psi_0^*(\vec{k},i\omega_n)
\right]
\label{psi0}
\end{equation}
where $\delta\psi_0({\vec k},i\omega_n)=\int d{\vec r} \int d\tau
e^{i\vec{k}\cdot \vec{r}}\delta\psi_0(\vec{r},\tau)$ and
$\delta\psi_0^*({\vec k},i\omega_n)=\int d{\vec r} \int d\tau
e^{i\vec{k}\cdot \vec{r}}\delta\psi_0^*(\vec{r},\tau)$.
Similarly, one finds
\begin{equation}
\delta\psi_0^*(\vec{k},i\omega_n)=
\frac{1}{-i\omega_n-\tilde{\varepsilon}_{\vec k}}
\left[
\frac{1}{\sqrt{\Omega}}\delta U(\vec{k},i\omega_n)
+ \frac{2g}{\sqrt{\Omega}} \delta \tilde{\rho}(\vec{k},i\omega_n)
+ g\rho_0 \delta\psi_0(\vec{k},i\omega_n)
\right]
\label{psi0star}
\end{equation}
Equations\ (\ref{psi0}) and (\ref{psi0star}) may substituted into
the definition of $\delta\rho_0$ to obtain
\begin{equation}
\left[ \frac{(i\omega_n)^2}{\tilde{\varepsilon}_{\vec k}-g\rho_0}
-\tilde{\varepsilon}_{\vec k}-g\rho_0 \right] \delta
\rho_0(\vec{k},i\omega_n) -4g\rho_0
\delta\tilde{\rho}(\vec{k},i\omega_n) =2g\rho_0\delta U
(\vec{k},i\omega_n), \label{second_eq}
\end{equation}
providing the second equation needed to compute the density disturbance.

Equations\ (\ref{first_eq}) and (\ref{second_eq}) may be combined
into a single matrix equation,
\begin{equation}
\label{matrix_eq} \left(\begin{array}{cc}
\frac{1}{\tilde{P}({\vec q},i\omega_n)}-2g &
-2g\\
-2g & \frac{1}{2\rho_0}\left[\frac{(i\omega_n)^2}
{\tilde{\varepsilon}_{\vec{q}}-g\rho_0}
-(\tilde{\varepsilon}_q+g\rho_0)\right]
\end{array}\right)
\left(\begin{array}{c}
\delta\tilde{\rho}(\vec{q},i\omega_n) \\
\delta\rho_0(\vec{q},i\omega_n)\end{array}\right)
=\left(\begin{array}{c} 1\\
1\end{array}\right)\delta U(\vec{q},i\omega_n).
\end{equation}
Collective modes of the system propagate when $\delta\tilde{\rho}$,
$\delta\rho_0 \ne 0$ even if $\delta U=0$.  This is only possible if
the matrix on the left hand side of Eq.\ (\ref{matrix_eq}) has
vanishing determinant,
\begin{equation}
\left[\frac{1}{\tilde{P}(\vec{q},i\omega_n)}-2g\right]
\left[\frac{(i\omega_n)^2}{\tilde{\varepsilon}_{\vec{q}}-g\rho_0}
-(\tilde{\varepsilon}_{\vec{q}}+g\rho_0)\right] -8g^2\rho_0=0.
\label{eq:modes}
\end{equation}
Upon analytic continuation ($i\omega_n \rightarrow \omega + i\eta$),
Eq.\ (\ref{eq:modes}) supports a linearly dispersing gapless mode
(zero sound) for any $\rho_0 \ne 0$, as expected for an infinite
uniform superfluid. This is a non-trivial check that our formalism
obtains physically sensible results.  Before analyzing this in
further detail, we demonstrate that the same result may be obtained
directly from the linear response formalism.  While somewhat more
complex to carry through, this latter approach allows us to look at
very general response functions, and so yields more information than
the density responses above.

\subsection{Density-Density Response Function and Collective Excitations}

Our starting point for this analysis is Eq.\ (\ref{eq:chi-eq}). For
a homogenous system, $\chi^R$ vanishes unless the indices are such
that momentum conservation is respected.  Thus we may set
$\lambda=\vec{k}_1-\vec{q}/2$, $\mu=\vec{k}_1+\vec{q}/2$,
$\alpha=\vec{k}_2-\vec{q}/2$, and $\beta=\vec{k}_2+\vec{q}/2$.
Denoting
\begin{align}
\chi^R_{\vec{k}_1\vec{k}_2}(\vec{q};\omega)
&=\chi^R_{\vec{k}_1-\frac{\vec{q}}{2}\vec{k}_1+\frac{\vec{q}}{2},
\vec{k}_2-\frac{\vec{q}}{2}\vec{k}_2+\frac{\vec{q}}{2}}(\omega),
\end{align}
one obtains
\begin{align}
\label{eq:chi-homogenous-1}
\left[\omega+\varepsilon_{\vec{k}_1+\frac{\vec{q}}{2}}
-\varepsilon_{\vec{k}_1-\frac{\vec{q}}{2}} \right]
\chi^R_{\vec{k}_1\vec{k}_2}(\vec{q};\omega)
&=\left(n_{\vec{k}_1+\frac{\vec{q}}{2}}-n_{\vec{k}_1-\frac{\vec{q}}{2}}\right)
\delta_{\vec{k}_1\vec{k}_2}
+\left(n_{\vec{k}_1+\frac12\vec{q}}-n_{\vec{k}_1-\frac12\vec{q}}\right)
\frac{2g}{\Omega}\sum_{\vec{k}_3}
\chi^R_{\vec{k}_3\vec{k}_2}(\vec{q};\omega)
\nonumber\\
&-g\rho_0\left(\delta_{\vec{k}_1,-\frac{\vec{q}}{2}}
-\delta_{\vec{k}_1,\frac{\vec{q}}{2}}\right)\left[\chi^R_{\frac{\vec{q}}{2}\vec{k}_2}
(\vec{q};\omega)+\chi^R_{-\frac{\vec{q}}{2}\vec{k}_2}(\vec{q};\omega)\right]
\nonumber\\
&-g\rho_0\left(\delta_{\vec{k}_1,-\frac{\vec{q}}{2}}
-\delta_{\vec{k}_1,\frac{\vec{q}}{2}}\right)
\chi^R_{\vec{k}_1\vec{k}_2}(\vec{q};\omega).
\end{align}
\end{widetext}
Bringing the last term to the left side of the equation has the
interesting effect of canceling the gap in the single-particle
spectrum, since
\begin{equation}
\label{eq:gap-cancellation}
\varepsilon_{\vec{k}_1+\frac{\vec{q}}{2}}
-\varepsilon_{\vec{k}_1-\frac{\vec{q}}{2}}+g\rho_0\left(\delta_{\vec{k}_1,-\frac{\vec{q}}{2}}
-\delta_{\vec{k}_1,\frac{\vec{q}}{2}}\right)=
\varepsilon^0_{\vec{k}_1+\frac{\vec{q}}{2}}
-\varepsilon^0_{\vec{k}_1-\frac{\vec{q}}{2}},
\end{equation}
where $\varepsilon^0_{\vec{k}}=\frac{\hbar^2\vec{k}^2}{2m}$ is the
free single-particle energy. This cancellation is only possible
because our linear response equation was generated in a way that is
consistent with the self-energy used in the static Hartree-Fock
analysis\ \cite{Kadanoff62}.

\begin{widetext}
Using Eq.\ (\ref{eq:chi-homogenous-1}), one can obtain
\begin{align}
\label{eq:chi-c} \left[\chi^R_{\frac12\vec{q}\vec{k}_2}
(\vec{q};\omega)+\chi^R_{-\frac12\vec{q}\vec{k}_2}(\vec{q};\omega)\right]
&=\frac{\omega^2-(\varepsilon^0_{\vec{q}})^2}
{\omega^2-(\varepsilon^0_{\vec{q}})^2+2g\rho_0\varepsilon^0_{\vec{q}}}
\left[\frac{N_0-n_{\vec{q}}}{\omega-\varepsilon^0_{\vec{q}}}
\delta_{\vec{k}_2,-\frac{\vec{q}}{2}}-\frac{N_0-n_{\vec{q}}}
{\omega+\varepsilon^0_{\vec{q}}}
\delta_{\vec{k}_2,\frac12\vec{q}}\right]
\nonumber\\
&+\frac{2(N_0-n_{\vec{q}})\varepsilon^0_{\vec{q}}}
{\omega^2-(\varepsilon^0_{\vec{q}})^2+2g\rho_0\varepsilon^0_{\vec{q}}}
\frac{2g}{\Omega}\sum_{\vec{k}_3}\chi^R_{\vec{k}_3\vec{k}_2}(\vec{q};\omega).
\end{align}
Substituting this result back into Eq.\ (\ref{eq:chi-homogenous-1})
leads to
\begin{align}
\label{eq:chi-homogenous-4}
\chi^R_{\vec{k}_1\vec{k}_2}(\vec{q};\omega)
&=\frac{n_{\vec{k}_1+\frac12\vec{q}}-n_{\vec{k}_1-\frac12\vec{q}}}
{\omega+\varepsilon^0_{\vec{k}_1+\frac12\vec{q}}
-\varepsilon^0_{\vec{k}_1-\frac12\vec{q}}}
\delta_{\vec{k}_1\vec{k}_2}
\nonumber\\
&\ \ \
-\frac{g}{\Omega}\frac{N_0\delta_{\vec{k}_1,-\frac12\vec{q}}
-N_0\delta_{\vec{k}_1,\frac12\vec{q}}}
{\omega+\varepsilon^0_{\vec{k}_1+\frac12\vec{q}}
-\varepsilon^0_{\vec{k}_1-\frac12\vec{q}}}
\frac{\omega^2-(\varepsilon^0_{\vec{q}})^2}
{\omega^2-(\varepsilon^0_{\vec{q}})^2+2g\rho_0\varepsilon^0_{\vec{q}}}
\left[\frac{N_0-n_{\vec{q}}}{\omega-\varepsilon^0_{\vec{q}}}
\delta_{\vec{k}_2,-\frac12\vec{q}}-\frac{N_0-n_{\vec{q}}}
{\omega+\varepsilon^0_{\vec{q}}}
\delta_{\vec{k}_2,\frac12\vec{q}}\right]
\nonumber\\
&\ \ \
+\left[\frac{n_{\vec{k}_1+\frac12\vec{q}}-n_{\vec{k}_1-\frac12\vec{q}}}
{\omega+\varepsilon^0_{\vec{k}_1+\frac12\vec{q}}-
\varepsilon^0_{\vec{k}_1-\frac12\vec{q}}}-
\frac{g}{\Omega}\frac{N_0\delta_{\vec{k}_1,-\frac12\vec{q}}
-N_0\delta_{\vec{k}_1,\frac12\vec{q}}}
{\omega+\varepsilon^0_{\vec{k}_1+\frac12\vec{q}}
-\varepsilon^0_{\vec{k}_1-\frac12\vec{q}}}
\frac{2(N_0-n_{\vec{q}})\varepsilon^0_{\vec{q}}}
{\omega^2-(\varepsilon^0_{\vec{q}})^2+2g\rho_0\varepsilon^0_{\vec{q}}}
\right]
\frac{2g}{\Omega}\sum_{\vec{k}_3}\chi^R_{\vec{k}_3\vec{k}_2}(\vec{q};\omega).
\end{align}
Now summing over $\vec{k}_1$ and $\vec{k}_2$, one gets
\begin{align}
\chi^R_{nn}(\vec{q};\omega)&=\tilde{P}(\vec{q};\omega)-\frac{2\rho_0\varepsilon^0_{\vec{q}}}
{\omega^2-(\varepsilon^0_{\vec{q}})^2+2g\rho_0\varepsilon^0_{\vec{q}}}
+2g\left[\tilde{P}(\vec{q};\omega)-\frac{2\rho_0\varepsilon^0_{\vec{q}}}
{\omega^2-(\varepsilon^0_{\vec{q}})^2+2g\rho_0\varepsilon^0_{\vec{q}}}\right]
\chi^R_{nn}(\vec{q};\omega).
\end{align}
\end{widetext}
One finally may express $\chi^R_{nn}(\vec{q};\omega)$ in a symmetric form,
\begin{align}
\label{eq:chi-nn}
\chi^R_{nn}(\vec{q};\omega)=\frac{\tilde{P}(\vec{q};\omega)+P_c(\vec{q};\omega)}
{1-2g\left[\tilde{P}(\vec{q};\omega)+P_c(\vec{q};\omega)\right]}
\end{align}
where we have denoted
\begin{equation}
P_c(\vec{q};i\omega)=\frac{2\rho_0\varepsilon^0_{\vec{q}}}
{\omega^2-(\varepsilon^0_{\vec{q}})^2+2g\rho_0\varepsilon^0_{\vec{q}}},
\end{equation}
which can be interpreted as the polarization function for the
condensate. The poles of $\chi_{nn}$ determine the collective
modes, and are given by solutions to
\begin{align}
\label{eq:pole-g-2}
1=2g\left[\tilde{P}(\vec{q};\omega)+P_c(\vec{q};i\omega)\right].
\end{align}
It is not difficult to show that this is identical to Eq.\
(\ref{eq:modes}).

\subsection{Discussion}

Several comments about the linear response
results as obtained from our TDHFA approach are in order.
Firstly, At $T=0$,
$\tilde{P}(\vec{q};\omega)=0$, and the density-density response function
becomes
\begin{align}
\label{eq:chi-nn-condensate}
\chi_{nn}(\vec{q};\omega,T=0)=\frac{P_c(\vec{q};\omega)}
{1-2gP_c(\vec{q};\omega)},
\end{align}
with the pole given by
\begin{align}
\omega^2(\vec{q},T=0)=2g\rho\varepsilon^0_{\vec{q}}+(\varepsilon^0_{\vec{q}})^2.
\end{align}
This is exactly the same as that obtained by linearizing the
time-dependent GP equation.

Secondly, we note that Eq.\ (\ref{eq:pole-g-2}) yields a propagating
gapless mode for {\it any} $\rho_0 \ne 0$.  This means our approach
correctly captures the superfluid mode of the system whenever it is
Bose-condensed. Thus in retaining the correct (gapped) structure for
the single-particle excitations, we see that the gapless superfluid
mode is not sacrificed. At $T>T_c$, $P_c(\vec{q};\omega)=0$, and
writing $\tilde{P}=P$, the density-density response function  is
\begin{align}
\label{eq:chi-nn-normal}
\chi_{nn}(\vec{q};\omega)=\frac{P(\vec{q};\omega)}
{1-2gP(\vec{q};\omega)},
\end{align}
with poles determined by
\begin{equation}
1=2gP(\vec{q};\omega),
\end{equation}
which is the known result for the normal Bose gas\
\cite{Fliesser01}.

\begin{figure}[t]
\includegraphics[width=1\columnwidth]{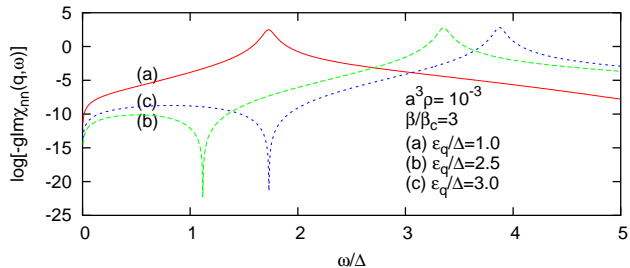}
\vspace{-5mm} \caption{(Color online)Imaginary part of the density
response for different values of $q$ as a function of $\omega$ for a
Bose condensate at inverse temperature $\beta$ (in units of the
inverse critical temperature $\beta_c$) with density $\rho$ and
scattering length  $a \equiv m/4\pi\hbar^2g$. } \label{fig1}
\end{figure}

Thirdly, because of the structure of Eqs.\ (\ref{matrix_eq}) and
(\ref{eq:pole-g-2}), one might expect to find two collective modes,
one in which the condensate and depleted particle densities
oscillate in phase, and the other in which they are out of phase.
These would correspond to zero and second sound, respectively. We
find however that for attainable values of $\mu(T)$ and $\rho_0(T)$
within the HFA, the two solutions of Eq.\ (\ref{eq:pole-g-2}) occur
with one at positive $\omega^2$ and the other at a negative value;
the second sound mode is thus overdamped. This is similar to the
weak coupling limit\ \cite{Griffin93}. However, unlike the weak
coupling limit, here the second sound is overdamped because of the
gap in the single-particle spectrum, which makes the depletion
particle density too small to support a propagating second sound
mode. The gap has another interesting effect on the interaction
between the condensate and depleted particles. This is illustrated
in Fig.\ \ref{fig1}, which depicts the imaginary part of the density
response function, $\chi_{nn}(q,\omega+i\eta)$ for fixed $q$ as a
function of $\omega$. Two features are prominently visible: a peak
(which becomes a delta function in the limit $q \rightarrow 0$)
representing the superfluid mode, and a zero in the response that
appears for $q^2/2m > 2\Delta \equiv 2g\rho_0$. The width of the
superfluid mode arises because of the interaction of the condensate
with the depletion particles, and vanishes at low temperature as
$e^{-\Delta/k_BT}$. Note we can obtain this temperature dependence
{\it only} because our choice of self-energy creates a gap between
the condensate and excited states, even though we use a single
Hamiltonian to describe them.  The zero arises when
$\omega^2=\bigl(\frac{q^2}{2m}\bigr)
\bigl(\frac{q^2}{2m}-2\Delta\bigr)$, and represents a frequency at
which $\delta\tilde{\rho}$ in Eq. \ref{matrix_eq} vanishes.  In this
situation the condensate precisely screens the perturbing potential
$\delta U$ for the depletion particles.  Since the depleted
particles are unperturbed and the condensate cannot absorb energy
away from the superfluid frequency, no energy can be absorbed by the
system, leading to the zero.  It is interesting to note that an
observation of this effect would allow one to measure the energy gap
of the system.

Finally, for small depletions $\tilde{\rho}$, the velocity of the
zero sound mode found from Eq.\ (\ref{eq:pole-g-2}) may be shifted
either upward or downward, depending on the numerical value of the
gas parameter.  This is most easily demonstrated by expanding the
equation for small $\tilde{\rho}$ and small $\tilde{P}$. The
correction to the sound velocity can then be shown to have the form
$\delta c_0 = \frac{1}{4mc_0}\lbrace g\Delta\rho_0+8g^2\rho
\tilde{P}_0 \rbrace$, where $c_0$ is the zero sound velocity at zero
temperature, $\Delta \rho_0$ the change in the condensate from its
zero temperature value, and $\tilde{P}_0 = \lim_{q \rightarrow 0}
\tilde{P}_{q}(\omega=c_0q)$.  For small values of $\rho$,
$\tilde{P}_0<0$, and the mode velocity decreases with temperature,
as is commonly found in the Bogoliubov approximation.  By contrast,
we find for larger values $\tilde{P}_0>0$, and for large enough
$\rho$, its value is sufficiently large to render $\delta c_0 > 0$.
This means relatively dense superfluids may have increasing zero
sound velocity with temperature, as is found in the RPA and appears
to be consistent with data for $^4$He\ \cite{Etters66}.  That the
TDHFA can capture both these limiting behaviors demonstrates the
utility of the method.

\section{Conclusion}
\label{sec:conclusion}

In this work, we demonstrated that the TDHFA may be developed in a
way that does not break the gauge symmetry of the Hamiltonian, thus
allowing us to correctly obtain a gapped single particle spectrum,
and yet correctly produces a gapless superfluid mode in the
collective response of an infinite uniform system.  By developing
both the depletion Green's function and response functions, we can
examine the single-particle and collective mode spectra separately.
The key to obtaining these features in a consistent way was to
retain the nonlocal terms in the self-energy (see Eqs.
(\ref{eq:static-HF-self-energy}) and (\ref{eq:TDHF-self-energy})),
which were required by the orthogonality of the single-particle
wavefunctions, and ultimately lead to a precise cancellation of the
gap in the single-particle energies when we calculate the density
response function (Eq.\ (\ref{eq:gap-cancellation})).

Finally, we point out that our equations for the infinite uniform
system (Eq.\ (\ref{matrix_eq})) are formally similar to equations
for the density response obtained using ensembles in which the gauge
symmetry is broken\ \cite{tosi}. This formal similarity however does
not generally occur; for example if we had studied a uniform but
finite size system (particles in a box), we would find further terms
that vanish in the thermodynamic limit, which do not appear in other
approaches. Generally speaking, in computing collective modes of a
Bose condensate, the challenge is to find how the condensate density
couples to the depleted particles in a way that is conserving\
\cite{Kadanoff62}, and preserves the gapless mode expected for an
infinite system\ \cite{Griffin96}. As discussed above, the TDHFA is
guaranteed to be conserving as it is controlled by a single
Hamiltonian (in contrast to many other methods), and we have
demonstrated that it correctly produces the expected superfluid
mode. Beyond the case we studied in detail in presenting our method
here, the TDHFA can be used to study inhomogeneous and/or finite
size systems, and may be very naturally generalized to handle
multicomponent Bose systems, rotating systems, and even
boson-fermion mixtures.  Future studies will focus on these
applications.

\acknowledgements

This work was supported by the NSF Grant No. DMR0454699. The authors
thank Allan Griffin and Wei Zhang for useful comments and
discussions.

\appendix
\section{Proof of Eq.\ (\ref{eq:vare-E})}
\label{app:delta-E} The off diagonal  matrix elements of $\Sigma$
can be found by varying the orthogonal basis $\psi_\alpha$.
We assume that $|\Psi\rangle$ is a Hartree-Fock
permanent constructed from orthonormal single-particle states
$\psi_\alpha$, which can be expressed as
\begin{align}
\label{eq:permanent}
|\Psi\rangle=\prod_{j}\frac{(a_{\alpha_j}^\dagger)^{n_{\alpha_j}}}{\sqrt{n_{\alpha_j}!}}
|0\rangle,
\end{align}
where $|0\rangle$ is a vacuum state. The variation of the
single-particle states may be written as
\begin{equation}
\label{eq:pert-A}
\delta\psi_\alpha=\sum_\gamma\eta_{\alpha\gamma}\psi_\gamma, \text{
or } \delta
a^\dagger_\alpha=\sum_\gamma\eta_{\alpha\gamma}a^\dagger_\gamma,
\end{equation}
with the orthonormality of the basis $\psi_\alpha$ implying that \cite{Huse82}
\begin{align}
\eta_{\alpha\beta}+\eta^*_{\beta\alpha}=0.
\end{align}
We can prove that the change of the permanent (\ref{eq:permanent})
due to the basis change (\ref{eq:pert-A}) can be generally expressed
as
\begin{align}
\label{eq:vare-Psi} |\delta\Psi\rangle=\sum_{\alpha\beta}
\eta_{\alpha\beta}a^\dagger_\beta a_\alpha|\Psi\rangle.
\end{align}
This can be shown by substituting Eq.\ (\ref{eq:pert-A}) into Eq.\
(\ref{eq:permanent}),
\begin{align}
|\delta\Psi\rangle&=\sum_{i,\beta} \eta_{\alpha_i\beta}\prod_{j\ne
i,\beta}\frac{(a_{\alpha_{j}}^\dagger)^{n_{\alpha_{j}}}}{\sqrt{n_{\alpha_j}!}}
(a_{\alpha_i}^\dagger)^{n_{\alpha_i}-1}
\frac{n_{\alpha_i}}{\sqrt{n_{\alpha_i}!}}
a^\dagger_\beta
|0\rangle.
\end{align}
Now moving $a^\dagger_\beta$ to the left side and noticing
$\frac{n_{\alpha_i}}{\sqrt{n_{\alpha_i}!}}
(a^\dagger_{\alpha_i})^{n_{\alpha_i}-1}|0\rangle=
\sqrt{n_{\alpha_i}}|n_{\alpha_i}-1\rangle=a_{\alpha_i}|n_{\alpha_i}\rangle$,
one obtains Eq.\ (\ref{eq:vare-Psi}).

The change of the free energy due to the basis change is
\begin{align}
\delta\Omega=\frac{\sum_i^\prime e^{-\beta(E_i-\mu N)}\left[
\langle\delta\Psi_i|H|\Psi_i\rangle+
\langle\Psi_i|H|\delta\Psi\rangle\right]} {\sum_i^\prime
e^{-\beta(E_i -\mu N)}},
\end{align}
where $E_i$ is the unperturbed Hartree-Fock energy corresponding to
the $\Psi_i$. Now substituting Eq.\ (\ref{eq:vare-Psi}) into above,
one obtains
\begin{align}
\delta\Omega&=\sum_{\mu\ne\nu}\eta_{\nu\mu} \frac{\sum_i
e^{-\beta(E_i-\mu N)}\langle \Psi_i|[H,a^\dagger_\mu
a_\nu]|\Psi_i\rangle} {\sum_ie^{-\beta( E_i-\mu)}}
\nonumber\\
&=\sum_{\mu\ne\nu}\eta_{\nu\mu} \langle[H,a_\mu^\dagger
a_\nu]\rangle,
\end{align}
which is the result of Eq.\ (\ref{eq:vare-E}).

\section{Perturbation Theory for Wavefunctions }
\label{sec:phi-perturbation}

In this Appendix, we seek a formal perturbation solution of the TDHF
equation since we need a relation of the functional derivative of
$W_\alpha$ with respect to $\delta U$.

If the time-dependent external field $\delta U(\vec{r}t)$ is weak,
we can solve the time-dependent single-particle wavefunctions
$\phi_\alpha(\vec{r}t)$ using time-dependent perturbation theory. In
order to do this, we expand the time-dependent single-particle
wavefunction $\phi_\alpha(\vec{r}t)$ in terms of the static HF basis
$\psi_\alpha(\vec{r})$ for times $t>t_0\rightarrow-\infty$ as
\begin{equation}
\phi_\alpha(\vec{r}t;\delta U)=\sum_\beta
C_{\alpha\beta}(t)e^{-i\varepsilon_\beta t}\psi_\beta(\vec{r}).
\end{equation}
We seek a solution for $C_{\alpha\beta}(t)$ up to the first order of
$\delta U(t)$. Substituting this into Eq.
(\ref{eq:HF-eq-T-finite}), after expanding the self-energy $\Sigma$
to first order of $\delta U(\vec{r}t)$ and using
Eq. (\ref{eq:static-HF-eq}), one
has
\begin{widetext}
\begin{align}
\label{eq:C-equation-1} \sum_\beta
i\frac{dC_{\alpha\beta}(t_1)}{dt_1} e^{-i\varepsilon_\beta
t_1}\psi_\beta(\vec{r}_1)&=\delta U(\vec{r}_1t_1)\sum_\beta
C_{\alpha\beta}(t_1)e^{-i\varepsilon_\beta
t_1}\psi_\beta(\vec{r}_1) \nonumber\\
&+\sum_\beta C_{\alpha\beta}(t_1)e^{-i\varepsilon_\beta t_1}
\int\!\!d\vec{r}_2dt_2d\vec{r}_3\int_{-\infty}^{t_1}
dt_3\left.\frac{\delta\Sigma(\vec{r}_1t_1,\vec{r}_2t_2,\delta
U)}{\delta U(\vec{r}_3t_3)}\right|_{\delta U=0}\delta
U(\vec{r}_3t_3)\psi_\beta(\vec{r}_2).
\end{align}
We can rewrite Eq. (\ref{eq:C-equation-1}) as
\begin{align}
\label{eq:C-equation} \sum_\beta i\frac{dC_{\alpha\beta}(t_1)}{dt_1}
e^{-i\varepsilon_\beta t_1}\psi_\beta(\vec{r}_1)&=\delta
U(\vec{r}_1t_1)\sum_\beta C_{\alpha\beta}(t_1)e^{-i\varepsilon_\beta
t_1}\psi_\beta(\vec{r}_1) \nonumber\\
&+\sum_\beta C_{\alpha\beta}(t_1)e^{-i\varepsilon_\beta t_1}
\int\!\!d\vec{r}_2d\vec{r}_3\int_{-\infty}^{\infty}
dt_3\Gamma^R(\vec{r}_1\vec{r}_3,\vec{r}_2\vec{r}_3,t_1-t_3)
\delta U(\vec{r}_3t_3)\psi_\beta(\vec{r}_2).
\end{align}
Multiplying both sides of Eq. (\ref{eq:C-equation}) by
$\psi^*_\gamma(\vec{r}_1)$, integrating over $\vec{r}_1$, and using
the Fourier expansion
\begin{equation}
\label{eq:U-expansion} \delta
U(\vec{r}t)=\int\!\!\frac{d\omega}{2\pi}\delta
U(\vec{r};\omega)e^{-i\omega t},
\end{equation}
one gets
\begin{align}
i\frac{dC_{\alpha\beta}(t_1)}{dt_1} &=
\sum_\gamma\int\!\frac{d\omega}{2\pi}
C_{\alpha\gamma}(t_1)e^{-i(\omega+\varepsilon_\gamma-\varepsilon_\beta)t_1}
\left[\langle\beta|\delta U(\omega)|\gamma\rangle
+\sum_{\lambda\mu}\Gamma^R_{\beta\gamma,\lambda\mu}(\omega)
\langle\lambda|\delta U(\omega)|\mu\rangle\right].
\end{align}
Writing
\begin{equation}
C_{\alpha\beta}(t)=\delta_{\alpha\beta}+C^{(1)}_{\alpha\beta}(t)
+\mathcal{O}(\delta U^2),
\end{equation}
one finds to first order
\begin{align}
i\frac{dC^{(1)}_{\alpha\beta}(t_1)}{dt_1} &=
\int\!\frac{d\omega}{2\pi}
e^{-i(\omega+\varepsilon_\alpha-\varepsilon_\beta)t_1}
\left[\langle\beta|\delta U(\omega)|\alpha\rangle
+\sum_{\lambda\mu}\Gamma^R_{\beta\alpha,\lambda\mu}(\omega)
\langle\lambda|\delta U(\omega)|\mu\rangle\right].
\end{align}
Integrating this equation and using the boundary condition
\begin{equation}
C^{(1)}_{\alpha\beta}(t\rightarrow-\infty)=0
\end{equation}
one gets
\begin{align}
C^{(1)}_{\alpha\beta}(t)=\int\!\frac{d\omega}{2\pi}
\frac{e^{-i(\omega+\varepsilon_\alpha-\varepsilon_\beta)t}}
{\omega+\varepsilon_\alpha-\varepsilon_\beta}
\left[\langle\beta|\delta U(\omega)|\alpha\rangle
+\sum_{\lambda\mu}\Gamma^R_{\beta\alpha,\lambda\mu}(\omega)
\langle\lambda|\delta U(\omega)|\mu\rangle\right].
\end{align}
We finally arrive at the time-dependent Hartree-Fock single-particle
wavefunctions up to the first order of $\delta U$,
\begin{align}
\label{eq:phi-perturbation} \phi_\alpha(\vec{r}t;\delta U)&=
e^{-i\varepsilon_\alpha t}
\left\{\psi_\alpha(\vec{r})+\sum_\beta\int\!\frac{d\omega}{2\pi}
\frac{e^{-i\omega t}} {\omega+\varepsilon_\alpha-\varepsilon_\beta}
\left[\langle\beta|\delta U(\omega)|\alpha\rangle
+\sum_{\lambda\mu}\Gamma^R_{\beta\alpha,\lambda\mu}(\omega)
\langle\lambda|\delta
U(\omega)|\mu\rangle\right]\psi_\beta(\vec{r})\right\},
\nonumber\\
\phi^*_\alpha(\vec{r}t;\delta U)&= e^{i\varepsilon_\alpha t}
\left\{\psi^*_\alpha(\vec{r})+\sum_\beta\int\!\frac{d\omega}{2\pi}
\frac{e^{i\omega t}} {\omega+\varepsilon_\alpha-\varepsilon_\beta}
\left[\langle\beta|\delta U(\omega)|\alpha\rangle^*
+\sum_{\lambda\mu}\Gamma^{R*}_{\beta\alpha,\lambda\mu}(\omega)
\langle\lambda|\delta
U(\omega)|\mu\rangle^*\right]\psi^*_\beta(\vec{r})\right\}.
\end{align}
The second terms of these equations are respectively the
variations $\delta\psi_{\alpha}({\vec r},t)$ and
$\delta\psi^*_{\alpha}({\vec r},t)$.
\end{widetext}

\bibliography{bec-zf2}

\begin{thebibliography}{19}
\expandafter\ifx\csname natexlab\endcsname\relax\def\natexlab#1{#1}\fi
\expandafter\ifx\csname bibnamefont\endcsname\relax
  \def\bibnamefont#1{#1}\fi
\expandafter\ifx\csname bibfnamefont\endcsname\relax
  \def\bibfnamefont#1{#1}\fi
\expandafter\ifx\csname citenamefont\endcsname\relax
  \def\citenamefont#1{#1}\fi
\expandafter\ifx\csname url\endcsname\relax
  \def\url#1{\texttt{#1}}\fi
\expandafter\ifx\csname urlprefix\endcsname\relax\def\urlprefix{URL }\fi
\providecommand{\bibinfo}[2]{#2}
\providecommand{\eprint}[2][]{\url{#2}}

\bibitem[{\citenamefont{Blaizot and Ripka}(1986)}]{Blaizot86}
\bibinfo{author}{\bibfnamefont{J.-P.} \bibnamefont{Blaizot}} \bibnamefont{and}
  \bibinfo{author}{\bibfnamefont{G.}~\bibnamefont{Ripka}},
  \emph{\bibinfo{title}{Quantum Theory of Finite Systems}}
  (\bibinfo{publisher}{The MIT Press}, \bibinfo{address}{London, England},
  \bibinfo{year}{1986}).

\bibitem[{gio()}]{giorgini}
\bibinfo{note}{A time-dependent Hartree-Fock-{\it Bogoliubov} approximation has
  been developed [S. Giorgini, Phys. Rev. A {\bf 61}, 063615 (2000)], but the
  structure of this is considerably different than what we describe below}.

\bibitem[{old()}]{old}
\bibinfo{note}{S.T. Beliaev, Sov. Phys. JETP {\bf 7}, 289 (1958); P. Hohenberg
  and P.C. Martin, Ann. Phys. {\bf 34}, 291 (1965); T.H. Cheung and A. Griffin,
  Phys. Rev. A {\bf 4}, 237 (1971)}.

\bibitem[{\citenamefont{Pethick and Smith}(2002)}]{Pethick02}
\bibinfo{author}{\bibfnamefont{C.~J.} \bibnamefont{Pethick}} \bibnamefont{and}
  \bibinfo{author}{\bibfnamefont{H.}~\bibnamefont{Smith}},
  \emph{\bibinfo{title}{Bose-Einstein Condensation in Dilute Gases}}
  (\bibinfo{publisher}{Cambridge University Press},
  \bibinfo{address}{Cambridge}, \bibinfo{year}{2002}).

\bibitem[{Gri()}]{Griffin96}
\bibinfo{note}{A. Griffin, Phys. Rev. B {\bf 53}, 9341 (1996); see also chapter
  by A. Griffin in M. Ignuscio, S. Stringari, and C. Wieman, {\it Bose-Einstein
  Condensation in Atomic Gases,} (IOS Press, Amsterdam, 1999)}.

\bibitem[{tos()}]{tosi}
\bibinfo{note}{A. Minguzzi an M.P. Tosi, J. Phys: Condens. Matter {\bf 9},
  10211 (1997); X.J. Liu et al., Phys. Rev. A {\bf 69}, 043605 (2004)}.

\bibitem[{\citenamefont{Griffin}(1993)}]{Griffin93}
\bibinfo{author}{\bibfnamefont{A.}~\bibnamefont{Griffin}},
  \emph{\bibinfo{title}{Excitations in a Bose-Condensed Liquid}}
  (\bibinfo{publisher}{Cambridge University Press},
  \bibinfo{address}{Cambridge}, \bibinfo{year}{1993}).

\bibitem[{die()}]{dielectric}
\bibinfo{note}{J. Reidl et al., Phys. Rev. A {\bf 61}, 043606 (2000); M.
  Flieser et al., Phys. Rev. A {\bf 64}, 013609 (2001)}.

\bibitem[{fix()}]{fix_n}
\bibinfo{note}{Y. Castin and R. Dum, Phys. Rev. A {\bf 57}, 3008 (1998); C.W.
  Gardiner and P. Zoller, Phys. Rev. A {\bf 61}, 033601 (2000); S.A. Morgan,
  Phys. Rev. A {\bf 69}, 023609 (2004).}

\bibitem[{\citenamefont{Kadanoff and Baym}(1962)}]{Kadanoff62}
\bibinfo{author}{\bibfnamefont{L.~P.} \bibnamefont{Kadanoff}} \bibnamefont{and}
  \bibinfo{author}{\bibfnamefont{G.}~\bibnamefont{Baym}},
  \emph{\bibinfo{title}{Quantum Statistical Mechanics}} (\bibinfo{publisher}{W.
  A. Benjamin, Inc}, \bibinfo{address}{New York}, \bibinfo{year}{1962}).

\bibitem[{\citenamefont{Huse and Siggia}(1982)}]{Huse82}
\bibinfo{author}{\bibfnamefont{D.~A.} \bibnamefont{Huse}} \bibnamefont{and}
  \bibinfo{author}{\bibfnamefont{E.~D.} \bibnamefont{Siggia}},
  \bibinfo{journal}{Journal of Low Temperature Phys.}
  \textbf{\bibinfo{volume}{46}}, \bibinfo{pages}{137} (\bibinfo{year}{1982}).

\bibitem[{oth()}]{others}
\bibinfo{note}{See for example V.V. Goldman, I.V. Silvera, and A.J. Leggett,
  Phys. Rev. B 24, 2870 (1981); V. Bagnato, D.E. Pritchard, and D. Kleppner,
  Phys. Rev. A {\bf 35}, 4354 (1987); J. Olive, Phys. Rev. B {\bf 39}, 4197
  (1989); M. Houbier and H.T.C. Stoof, Phys.Rev. A {\bf 54}, 5055 (1996); P.
  \"Ohberg and S. Stenholm, J. Phys. B {\bf 30}, 2749 (1997)}.

\bibitem[{dep()}]{depletion}
\bibinfo{note}{We use ``depletion'' rather than ``quasiparticle'' because the
  latter commonly refers to phonons of Bose-condensed systems}.

\bibitem[{\citenamefont{Ivanchemko and Lisyansky}(1995)}]{Iva95}
\bibinfo{author}{\bibfnamefont{Y.~M.} \bibnamefont{Ivanchemko}}
  \bibnamefont{and} \bibinfo{author}{\bibfnamefont{A.~A.}
  \bibnamefont{Lisyansky}}, \emph{\bibinfo{title}{Physics of Critical
  Fluctuations}} (\bibinfo{publisher}{Springer}, \bibinfo{address}{New York},
  \bibinfo{year}{1995}).

\bibitem[{\citenamefont{Leggett}(2001)}]{Leggett01}
\bibinfo{author}{\bibfnamefont{A.~J.} \bibnamefont{Leggett}},
  \bibinfo{journal}{Rev. Mod. Phys.} \textbf{\bibinfo{volume}{73}},
  \bibinfo{pages}{307} (\bibinfo{year}{2001}).

\bibitem[{\citenamefont{Kolomeisky et~al.}(2004)\citenamefont{Kolomeisky,
  Straley, and Kalas}}]{kolomeisky}
\bibinfo{author}{\bibfnamefont{E.}~\bibnamefont{Kolomeisky}},
  \bibinfo{author}{\bibfnamefont{J.}~\bibnamefont{Straley}}, \bibnamefont{and}
  \bibinfo{author}{\bibfnamefont{R.}~\bibnamefont{Kalas}},
  \bibinfo{journal}{Phys. Rev. A} \textbf{\bibinfo{volume}{69}},
  \bibinfo{pages}{063401} (\bibinfo{year}{2004}).

\bibitem[{\citenamefont{Mahan}(2000)}]{Mahan}
\bibinfo{author}{\bibfnamefont{G.}~\bibnamefont{Mahan}},
  \emph{\bibinfo{title}{Many-Particle Physics, 3rd Ed.}}
  (\bibinfo{publisher}{Plenum Publishers}, \bibinfo{address}{New York},
  \bibinfo{year}{2000}).

\bibitem[{\citenamefont{Fliesser et~al.}(2001)\citenamefont{Fliesser, Reidl,
  Sz\'epfalusy, and Graham}}]{Fliesser01}
\bibinfo{author}{\bibfnamefont{M.}~\bibnamefont{Fliesser}},
  \bibinfo{author}{\bibfnamefont{J.}~\bibnamefont{Reidl}},
  \bibinfo{author}{\bibfnamefont{P.}~\bibnamefont{Sz\'epfalusy}},
  \bibnamefont{and} \bibinfo{author}{\bibfnamefont{R.}~\bibnamefont{Graham}},
  \bibinfo{journal}{Phys. Rev. A} \textbf{\bibinfo{volume}{64}},
  \bibinfo{pages}{013609} (\bibinfo{year}{2001}).

\bibitem[{\citenamefont{Etters}(1966)}]{Etters66}
\bibinfo{author}{\bibfnamefont{R.~D.} \bibnamefont{Etters}},
  \bibinfo{journal}{Phys. Rev. Lett.} \textbf{\bibinfo{volume}{16}},
  \bibinfo{pages}{119} (\bibinfo{year}{1966}).

\end{thebibliography}

\end{document}